\documentclass[prd,aps,floats,preprintnumbers,onecolumn]{revtex4}
\usepackage{graphicx, epsfig}
\usepackage{amsmath}
\usepackage{amssymb}
\usepackage{psfig}
\usepackage{natbib}

\begin{document}

\title{Cosmological parameter estimation with large scale structure and supernovae data}

\author{Carolina J. Odman$^{\flat,\sharp}$, Mike Hobson$^\sharp$, Anthony Lasenby$^\sharp$ and Alessandro Melchiorri$^\flat$, }

\affiliation{ $^\flat$ Physics Department, University of Rome ``La Sapienza'', Ple Aldo Moro 2, 00185, Rome, Italy\\
$^\sharp$ Astrophysics Group, Cavendish Laboratory, Cambridge University, Cambridge, U.K.\\}

\begin{abstract}
Most cosmological parameter estimations are based on the same set of observations and are therefore not independent. Here, we test the consistency of parameter estimations using a combination of large-scale structure and supernovae data, without cosmic microwave background (CMB) data. 
We combine observations from the IRAS 1.2 Jy and Las Campanas redshift surveys, galaxy peculiar velocities and measurements of type Ia supernovae to obtain $h=0.57_{-0.14}^{+0.15}$, $\Omega_m=0.28\pm0.05$ and $\sigma_8=0.87_{-0.05}^{+0.04}$ in agreement with the constraints from observations of the CMB anisotropies by the WMAP satellite. We also compare results from different subsets of data in order to investigate the effect of priors and residual errors in the data. We find that some parameters are consistently well constrained whereas others are consistently ill-determined, or even yield poorly consistent results, thereby illustrating the importance of priors and data contributions. 
\end{abstract}

\maketitle


\section{Introduction}
The measurements of the cosmic microwave background (CMB) from the WMAP satellite \cite{Bennett03} and large-scale galaxy clustering from the 2dF \cite{2df_percival} and SLOAN \cite{pope_sdss} surveys have provided tight constraints on the value of  cosmological parameters \cite{tegmark03}. The combination of  observations of different nature has the power to tighten constraints on the parameters of cosmological models (see e.g. \cite{wmap-spergel}, \cite{tegmark02}, \cite{cjo}, \cite{slosar}, \cite{darkenergy}).
A shortcoming of this approach, however, is that most parameter estimations use the same data. They are not independent and their agreement comes as no surprise. Also, if one set of data suffers from uncontrolled systematic errors, those errors might affect the results and go unnoticed. It is therefore useful to check those results with an independent set of observations.

Some cosmological parameters such as the Hubble parameter $h$ are ill-determined from one type of observations alone and a joint analysis is necessary in order to determine its value. If such an analysis is not available some prior value is often assumed, based on some dedicated observation. For example, the HST key project constraint on the value of the Hubble constant $H_0 = 0.72 \pm 0.08 h$ km s$^{-1}$Mpc$^{-1}$ often helps to tighten the determination of other parameter values \cite{bridle_science}. The robustness of such results might depend strongly on the prior.

In this brief communication we use cosmological data to estimate cosmological parameters and to provide an independent test of more recent analyses. We combine data from the IRAS 1.2 Jy \cite{IRAS} and Las Campanas \cite{LCRS-KL} redshift surveys, galaxy peculiar velocity measurements \cite{PecVel} and measurements of type Ia supernovae \cite{lasenby-sne,efstathiou-sne}. We do not use any cosmic microwave background data. We examine our results according to subsets of data. We focus on three parameters: the matter density $\Omega_m$, the Hubble parameter $h$ and the amplitude of mass density fluctuations $\sigma_8$.

Our paper is organized as follows. In the sections \ref{lss-data} and \ref{lss-method} we  describe the data and method used in the analysis. The results are presented in section \ref{lss-results} and conclusions are presented in section \ref{lss-conclusions}.


\section{Large-scale structure and Supernovae data}\label{lss-data}

The IRAS $1.2$ $\rm{Jy}$ data set \cite{EBWhite92} is an all-sky (87.6$\% 
$) infra-red survey of 5313 galaxies. The density of galaxies in three dimensional space is sensitive to the mass density fluctuation parameter $\sigma_8$ and the inferred matter power spectrum depends on the shape parameter $\Gamma$.  The likelihood function was calculated for flat universes within linear theory by \cite{IRAS} and assumes that light traces mass via a linear bias factor $\left(\delta\rho / \rho\right) = b_{\textrm{IRAS}} \left( \delta \rho / \rho\right)_m$. The likelihood is a function of the 3 parameters $\sigma_8$, $\beta_{\textrm{IRAS}}$ and $\Gamma$.

The Las Campanas Redshift Survey (LCRS) probes the same parameters as the IRAS survey but both the data and the likelihood function calculation differ in several ways. IRAS detected galaxies in the infra-red whereas the Las Campanas  experiment measured $\sim 26,000$ galaxies at optical wavelengths \cite{LCRS-original}.  A different morphological selection took place in each of the experiments such that their bias parameters are different. The LCRS probes a deeper range of redshifts in finger-shaped volumes consisting of a northern and a southern sample. A Karhunen-Lo\`eve (KL) eigenmode basis was constructed to correlate the measured redshifts as described in \cite{LCRS-KL} (see also \cite{SDSS-KL}). The data were compared to matter power spectra expressed in terms of the KL basis from numerically simulated open-CDM structure formation scenarios with different values of $\sigma_8$, $\beta_{LCRS}$ and $\Gamma$. 

Unlike redshift surveys which measure the distribution of visible matter, peculiar velocity measurements are a dynamical probe of the matter power spectrum. Within the assumption that galaxies trace the large-scale gravitation induced velocity field it is expected that velocities obey linear theory better than density due to the non-local character of the gravitational field. This likelihood function was calculated in \cite{PecVel} using the Mark III catalogue of more than 3000 galaxy velocities with no morphological selection (see \cite{mark3} and references therein). 

The SNe Ia likelihood function used is calculated in \cite{efstathiou-sne} and \cite{lasenby-sne} by maximising the likelihood function using 42 high redshift supernovae measured by the Supernova Cosmology Project \cite{SCP-99}. Note that SNe Ia measurements probe cosmological parameters through the luminosity distance.  Unlike redshift surveys, SNe Ia are independent of the structure formation scenario so the SNe Ia and LSS combination of data is a powerful tool for constraining cosmological models.


\section{Method}\label{lss-method}

In order to perform a joint analysis, one must ensure a basis set of independent parameters from which the other parameters can be calculated so the parameter space depends on the combination of data sets.

The IRAS likelihood function was computed under the assumption of a flat universe. Hence only flat models are considered when IRAS data are combined with, e.g. supernovae data even if the latter likelihood function allows more freedom in the parameter space. Similarly, the Las Campanas likelihood function was calculated by comparing realisations of the survey from cold dark matter structure formation simulations. Therefore no hot or warm dark matter models can be constrained with the Las Campanas likelihood function.

The assumptions of each data set are propagated throughout the analysis because they change the way the parameter space can be defined and how the remaining parameters are calculated from the independent ones. This conservative approach in which only models common to all used data sets can be constrained, has the drawback that the results cannot always be compared at face value. The limitation to flat universes corresponds to a slice of the larger parameter space allowing for open and closed universes. Results from the larger parameter space, marginalised over curvature, cannot be considered statistically equivalent to results for flat universes only. The priors on the considered models for each combination are summarised in table \ref{prior-table}.

If the likelihood functions are not evaluated beyond some value the nearest border value of the likelihood function is used. Although the IRAS likelihood function is incomplete in the $\sigma_8$ dimension, this choice of extrapolation has no consequences when using the full combination of data sets. The analysis is carried out with {\scshape Bayesys}, a commercial MCMC software package \cite{skilling}. 50,000 samples are obtained for each possible combination of data. Since the resulting number density of samples in parameter space is proportional to the joint likelihood, the samples are then counted into bins and those one-dimensional distributions are automatically marginalised over the remaining parameters.

\begin{table}[!bht]
\begin{center}
\begin{tabular}{|r|c|r|c|}\hline
L & CDM & I + S & Flat\\
I & Flat & P + S & --\\
P & --& L + I + P & CDM + Flat \\
S & --& L + I + S & CDM + Flat \\
L + I & CDM + Flat & L + P + S & CDM \\
L + P & CDM & I + P + S & Flat \\
L + S & CDM & L + I + P + S & CDM + Flat \\
I + P & Flat & & \\\hline
\end{tabular}
\end{center}
\caption{Prior for each combination of four data sets: The Las Campanas (L) and IRAS (I) redshift surveys, galaxy peculiar velocities (P) and distant supernovae (S) and associated priors. ``CDM'' means only cold dark matter models were used in the likelihood function calculation. ``Flat'' means that only models of flat universes were considered and the likelihood function was generated under that assumption.}
\label{prior-table}
\end{table}


\section{Results}\label{lss-results}
\begin{figure}[!bht]
\begin{center}
\begin{tabular}{ccccccccc}
\tabularnewline 
\includegraphics[%
scale=0.19,angle=270]{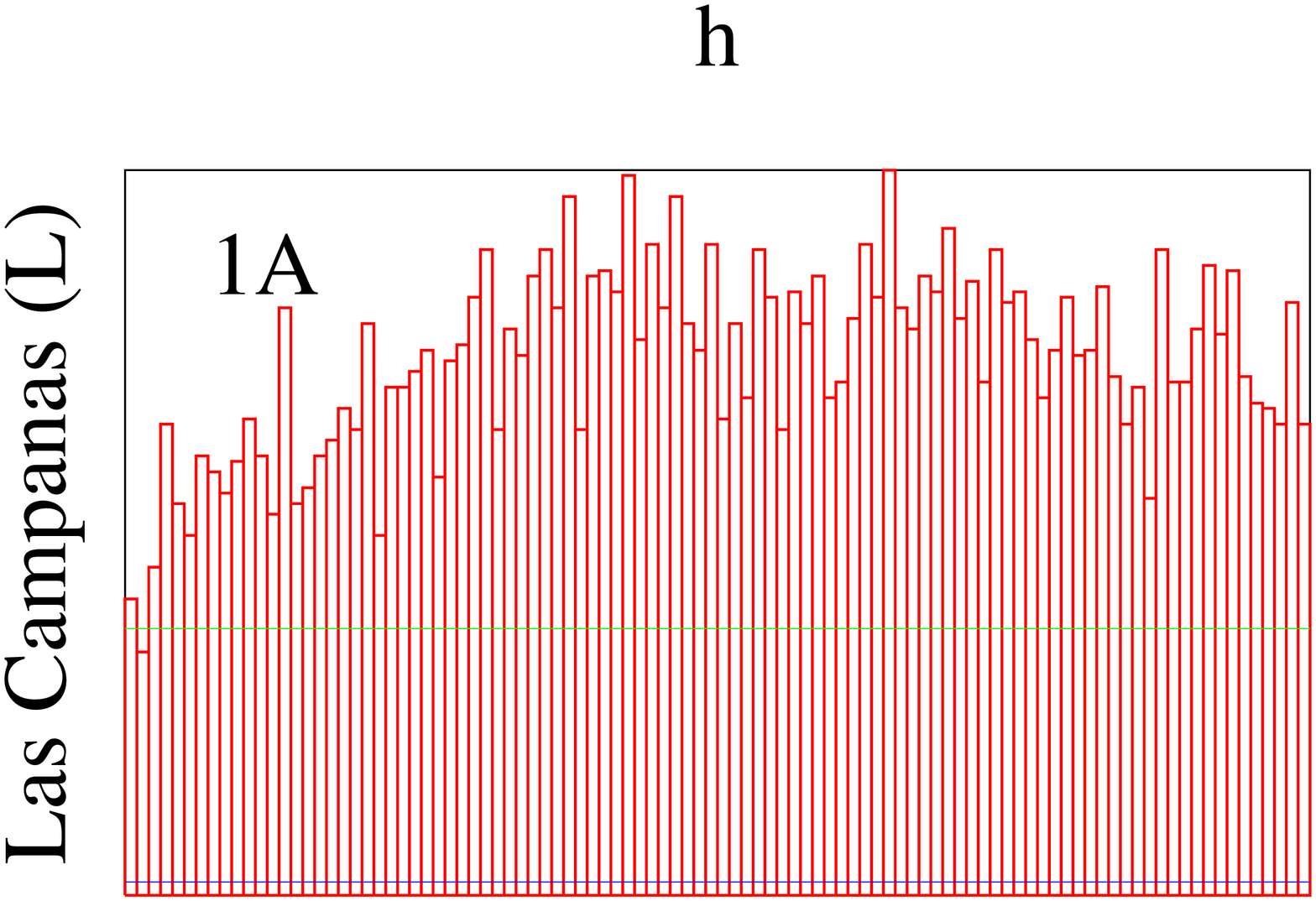}
\hspace{-6.5mm}
\includegraphics[%
scale=0.19,angle=270]{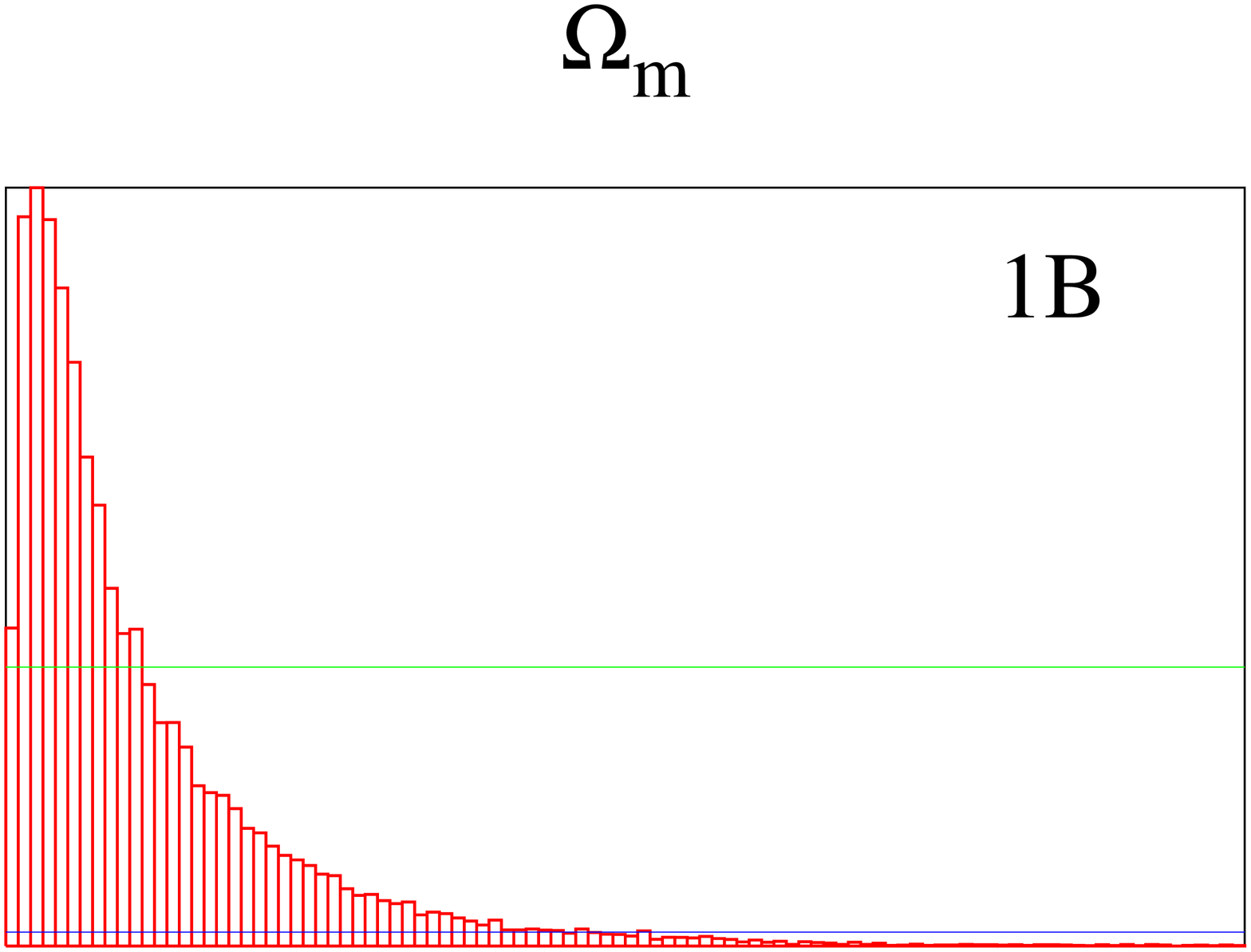}
\hspace{-6.5mm}
\includegraphics[%
scale=0.19,angle=270]{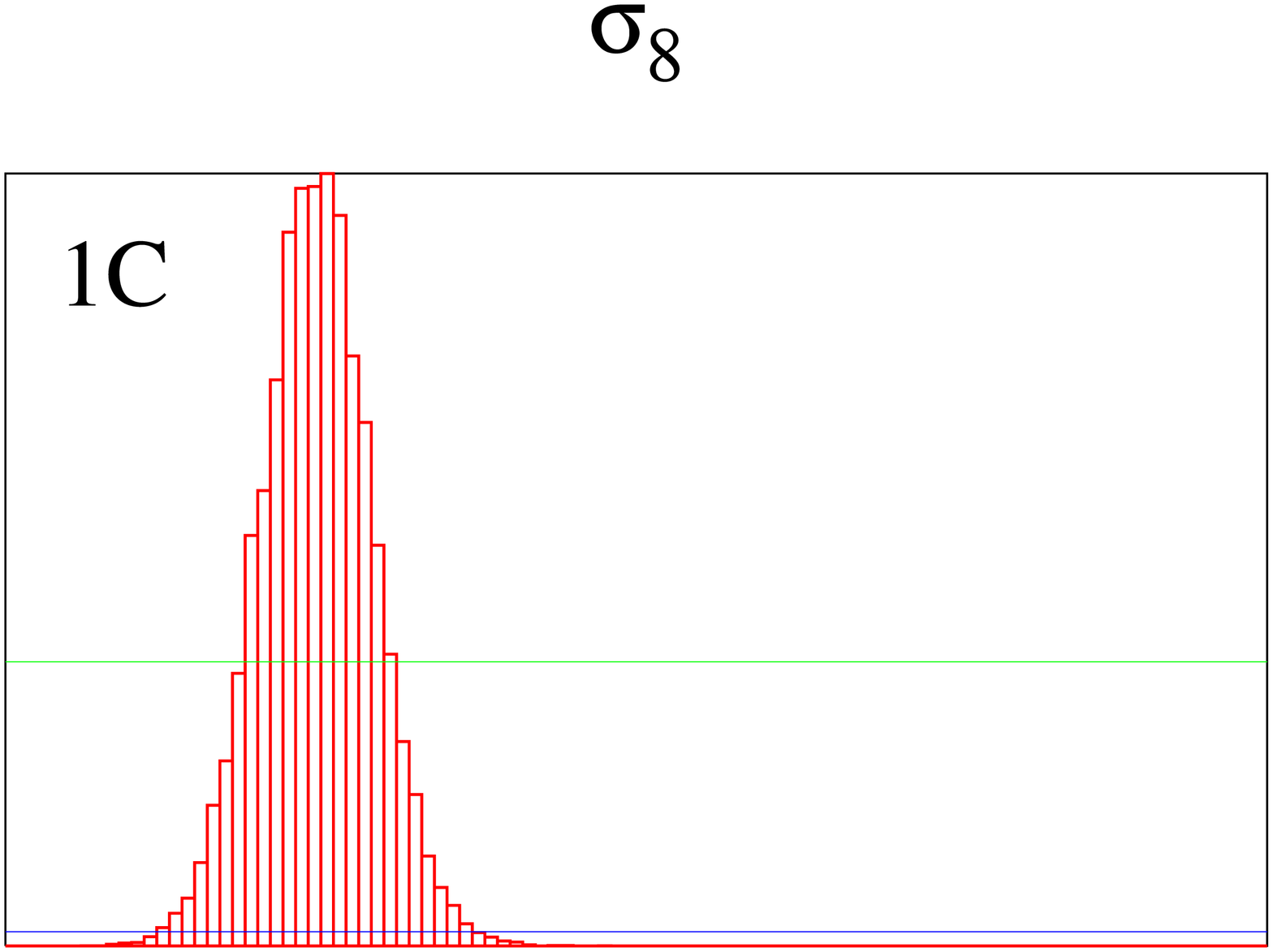}
\vspace{-9.8mm}
\tabularnewline
\includegraphics[%
scale=0.19,angle=270]{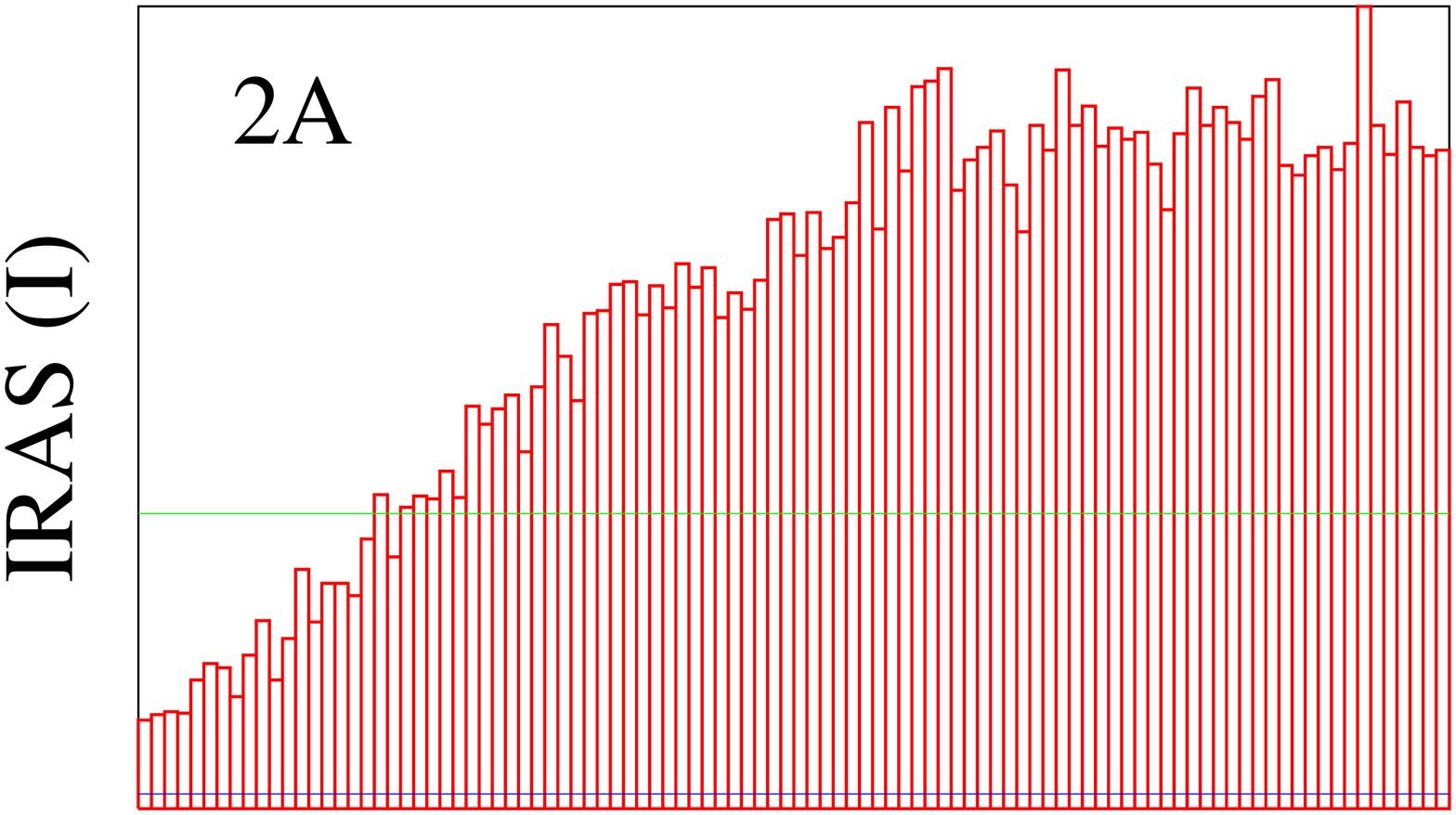}
\hspace{-6.5mm}
\includegraphics[%
scale=0.19,angle=270]{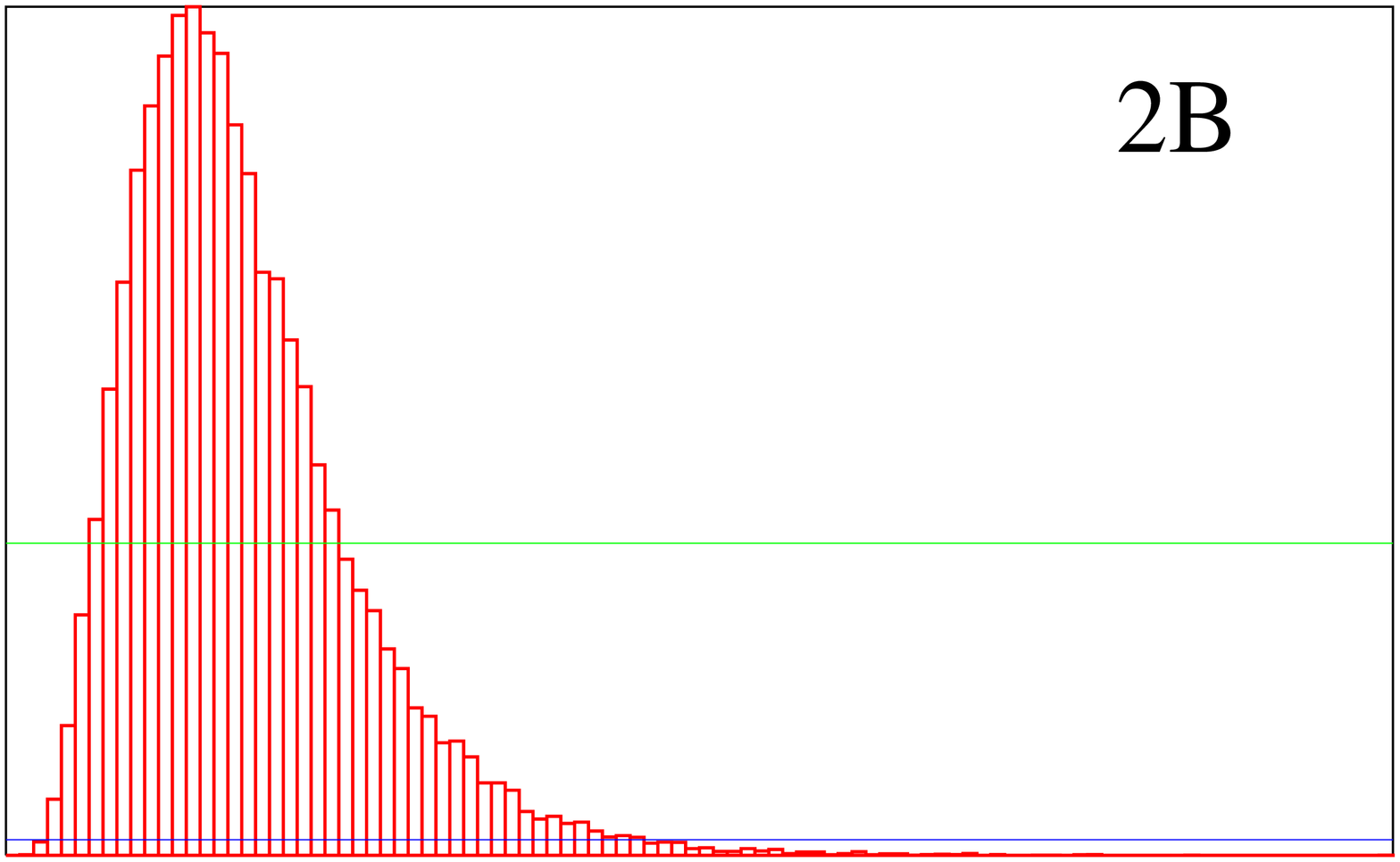}
\hspace{-6.5mm}
\includegraphics[%
scale=0.19,angle=270]{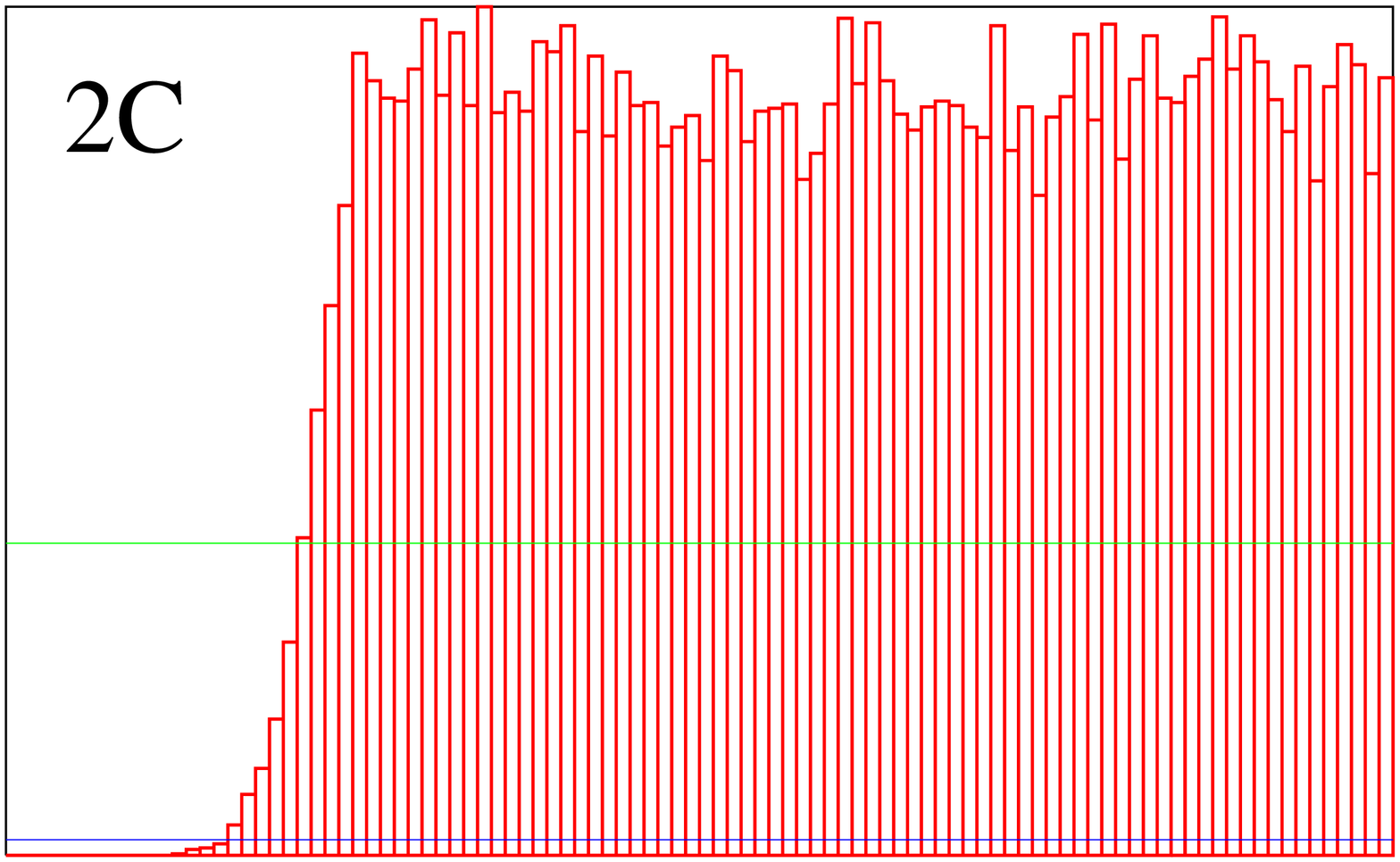}
\vspace{-9.8mm}
\tabularnewline
\includegraphics[%
scale=0.19,angle=270]{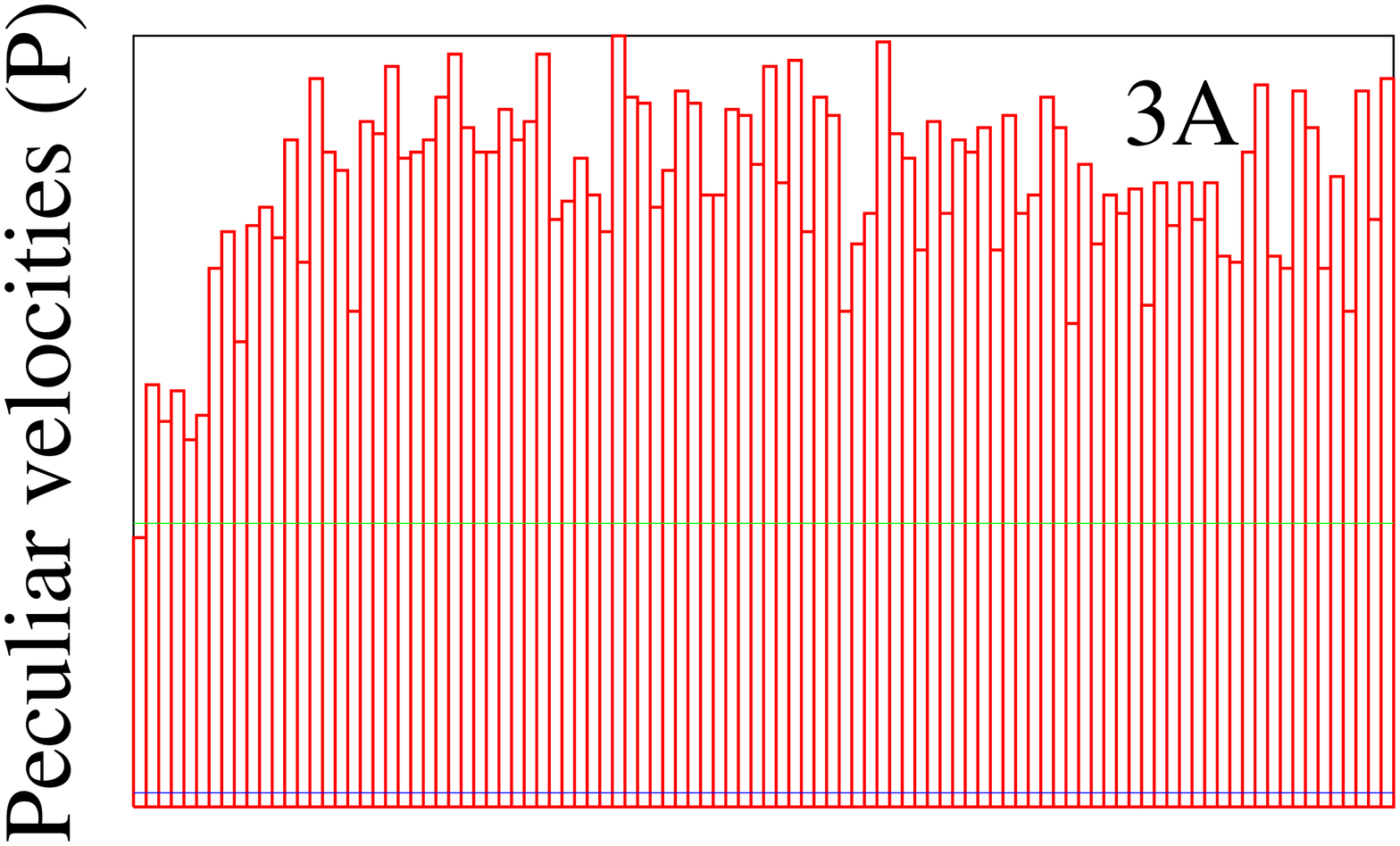}
\hspace{-6.5mm}
\includegraphics[%
scale=0.19,angle=270]{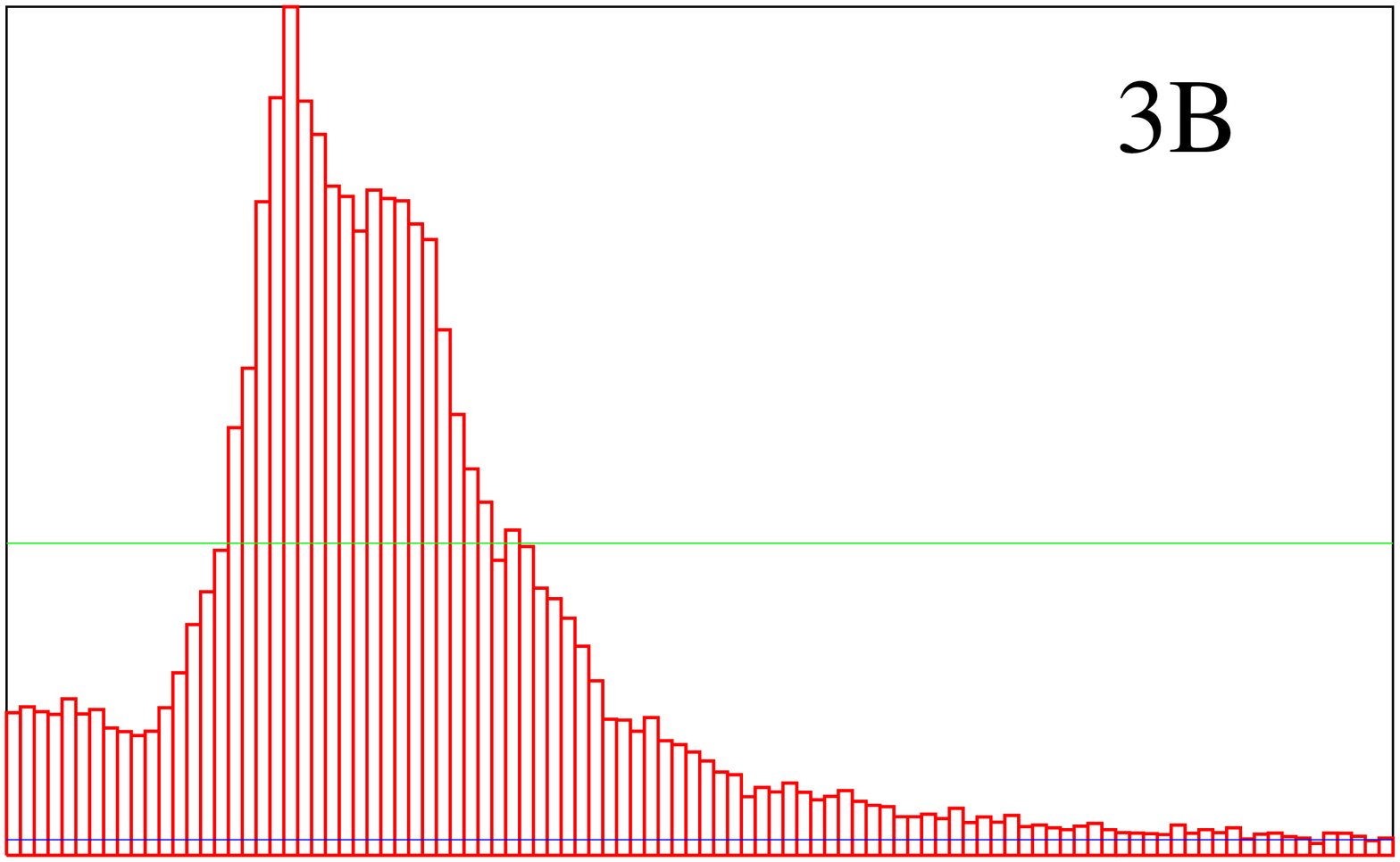}
\hspace{-6.5mm}
\includegraphics[%
scale=0.19,angle=270]{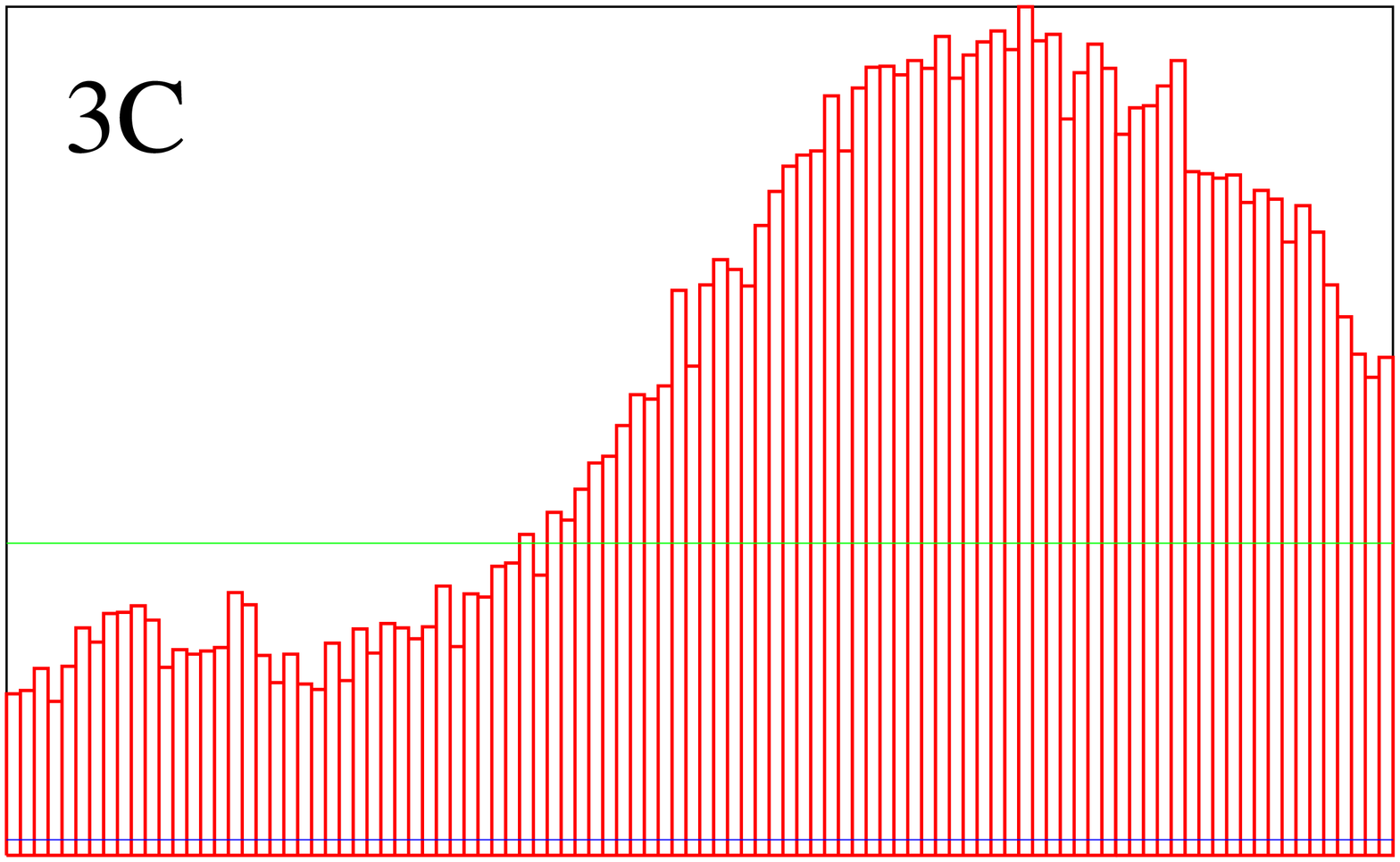}
\vspace{-9.8mm}
\tabularnewline
\includegraphics[%
scale=0.19,angle=270]{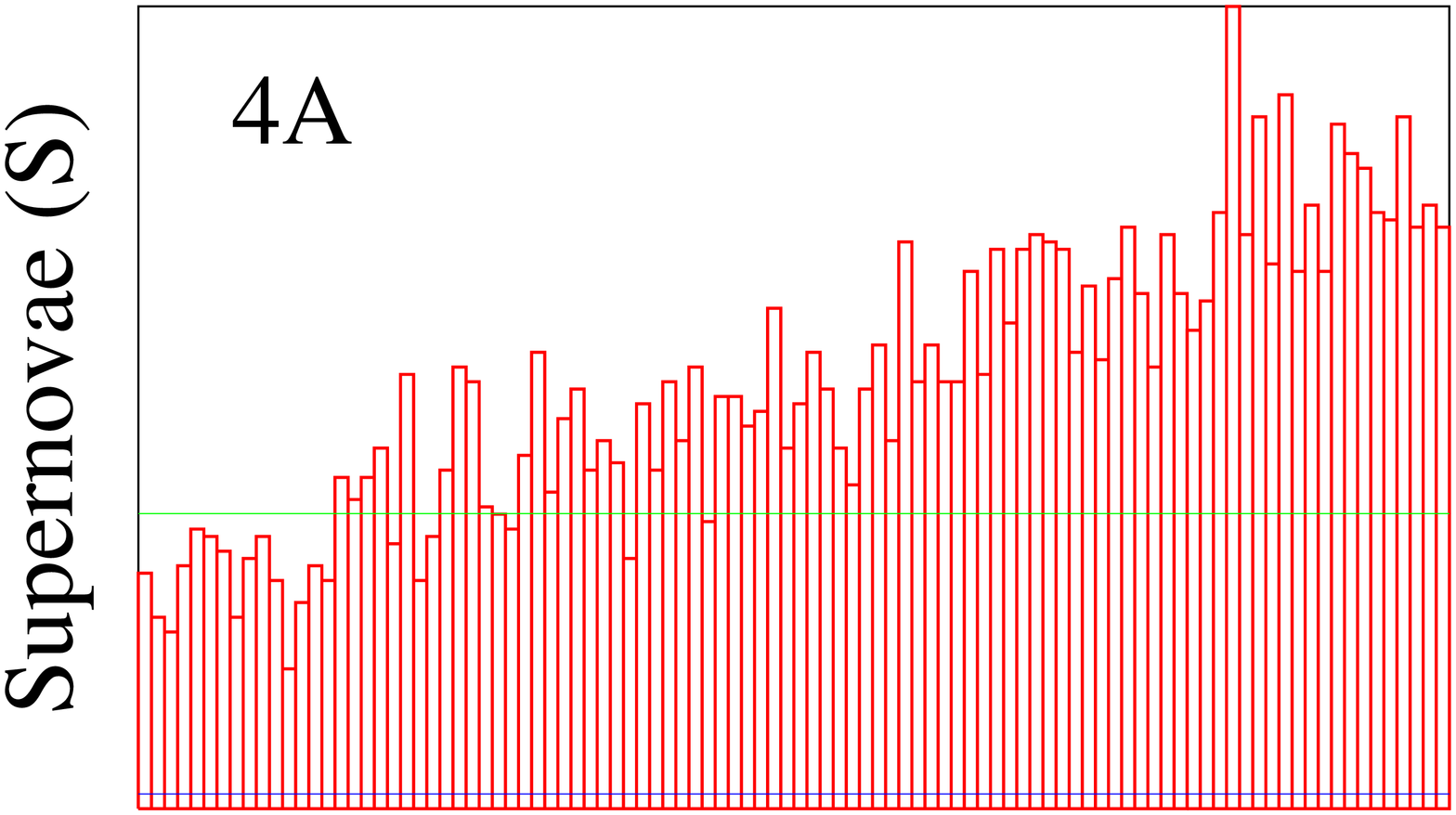}
\hspace{-6.5mm}
\includegraphics[%
scale=0.19,angle=270]{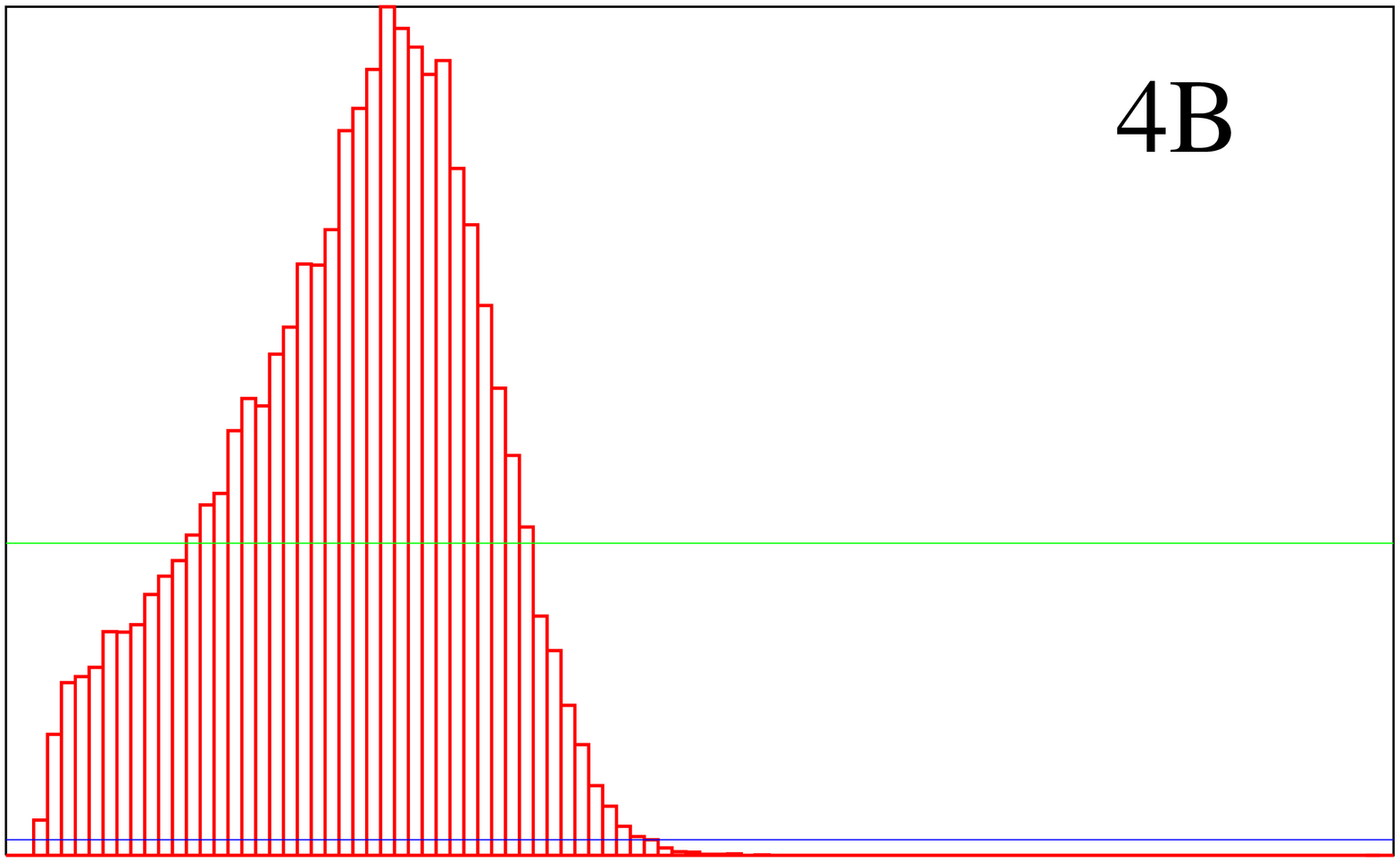}
\hspace{-6.5mm}
\includegraphics[%
scale=0.19,angle=270]{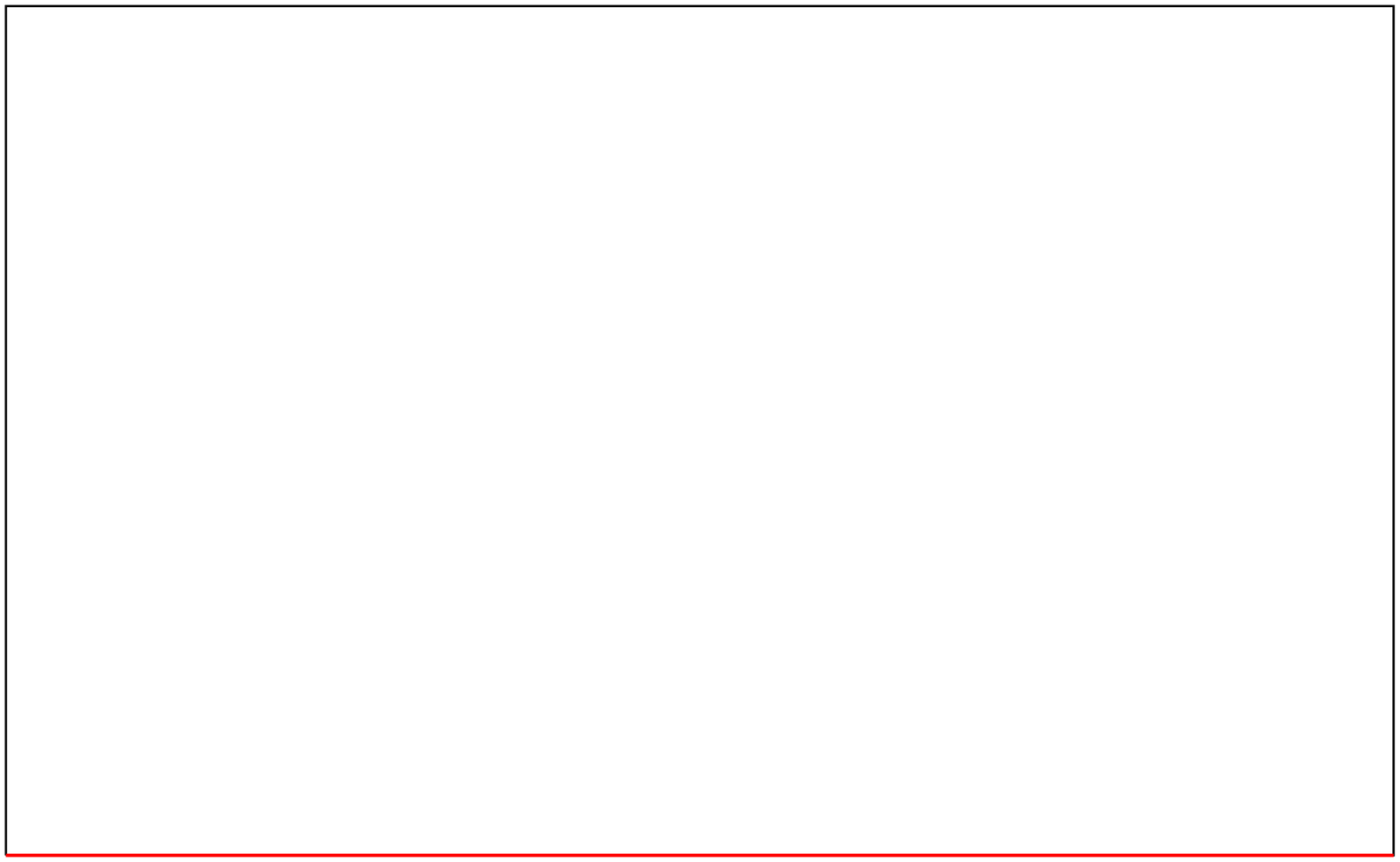}
\vspace{-9.8mm}
\tabularnewline
\includegraphics[%
scale=0.19,angle=270]{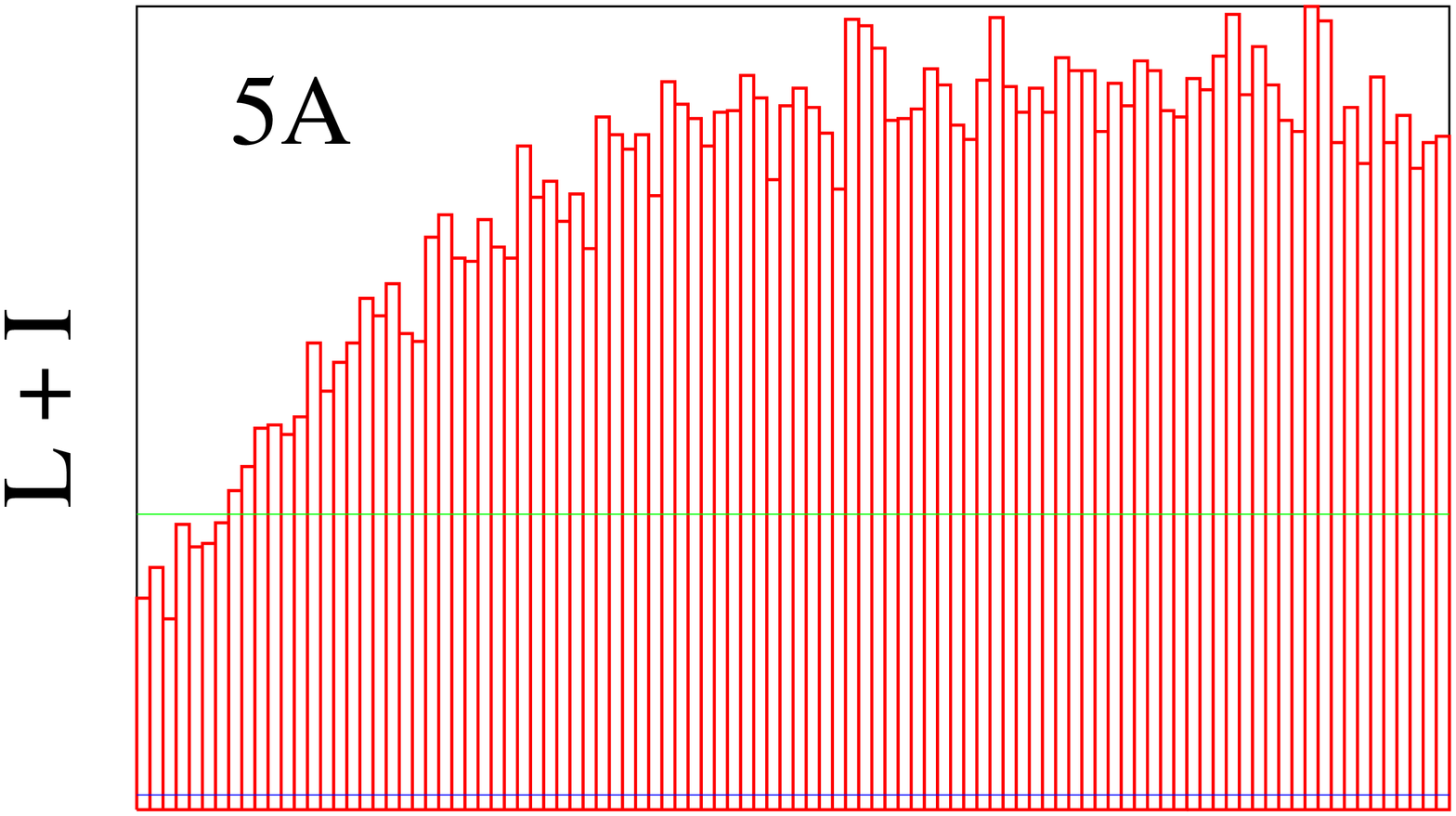}
\hspace{-6.5mm}
\includegraphics[%
scale=0.19,angle=270]{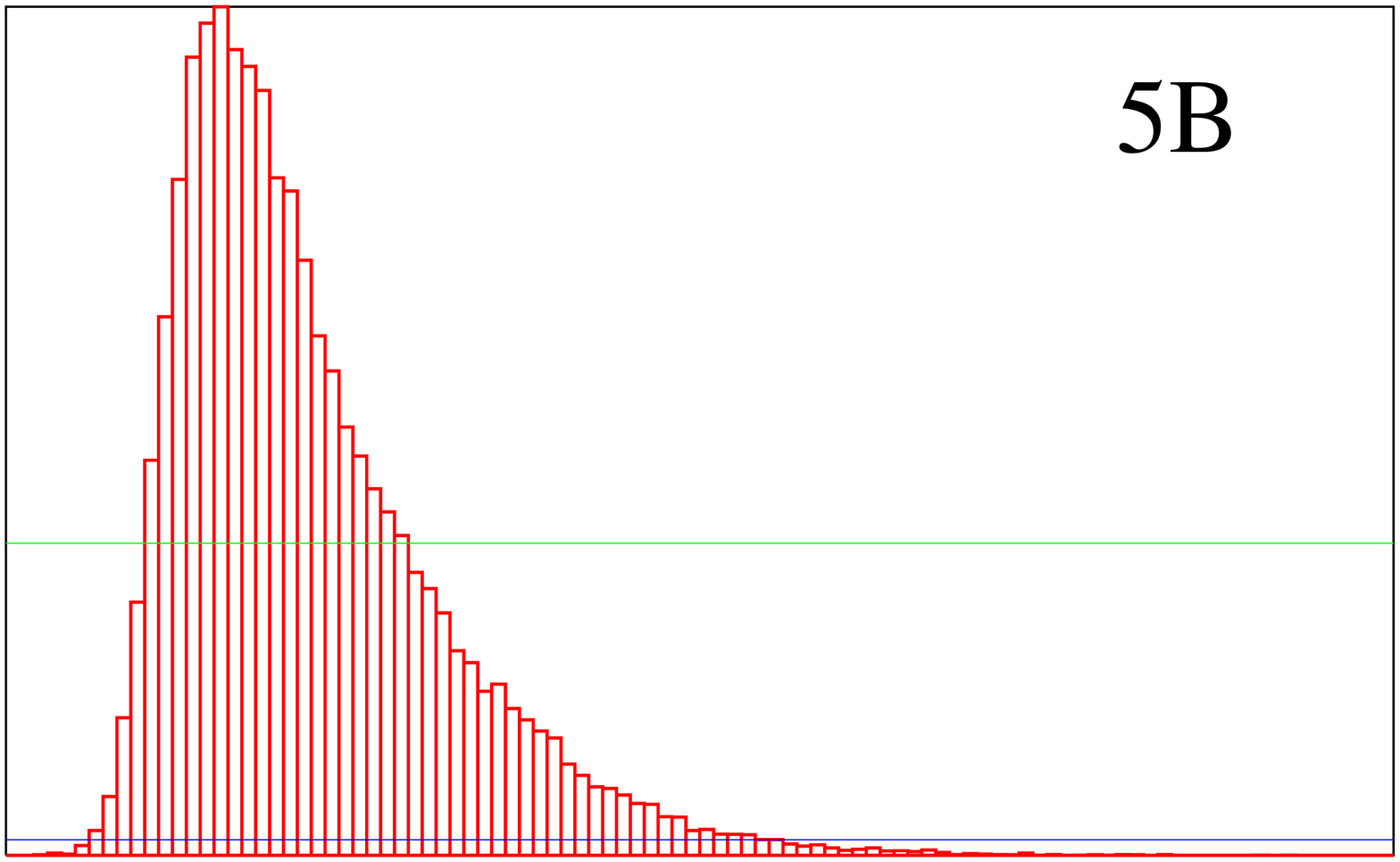}
\hspace{-6.5mm}
\includegraphics[%
scale=0.19,angle=270]{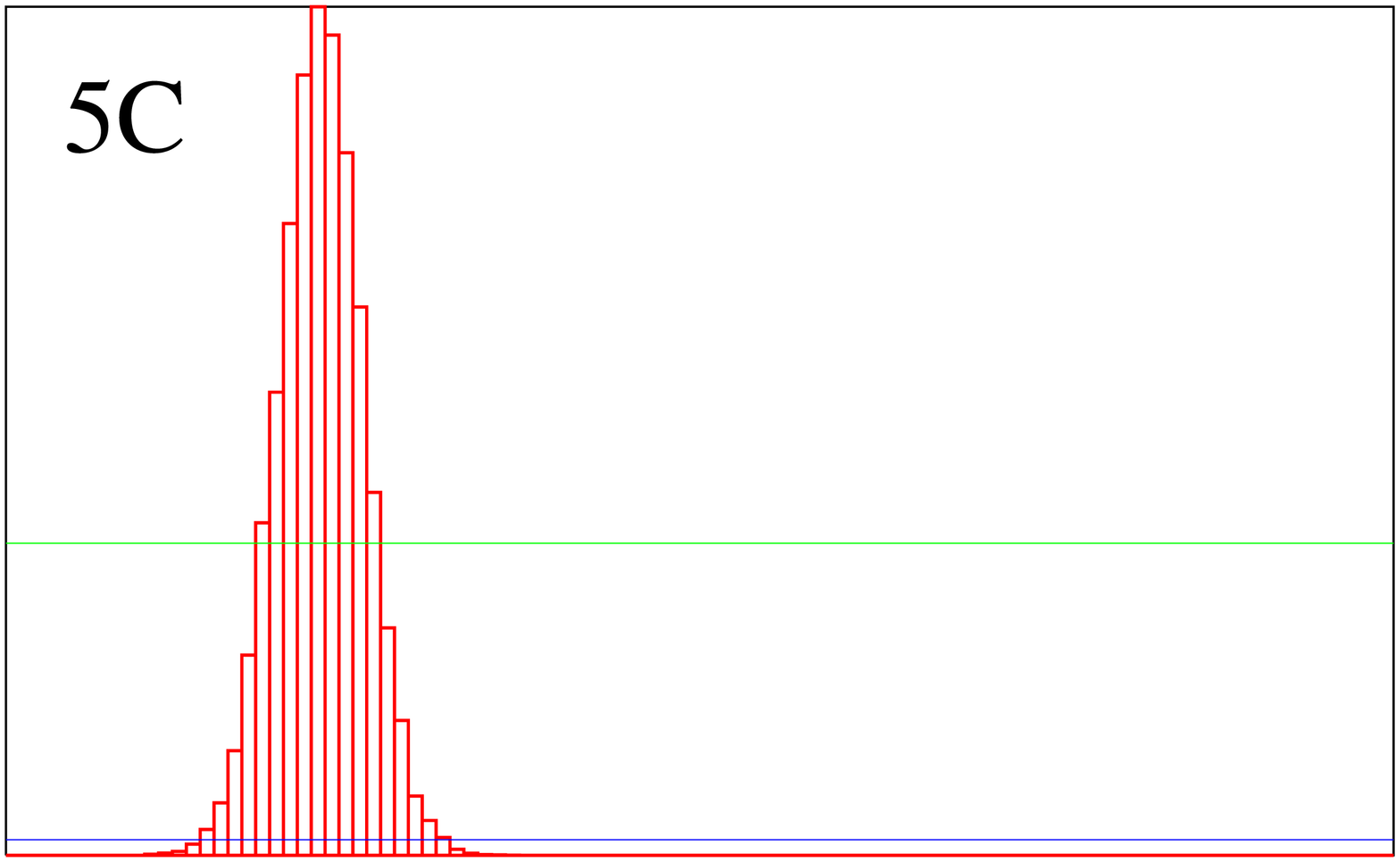}
\vspace{-9.8mm}
\tabularnewline
\includegraphics[%
scale=0.19,angle=270]{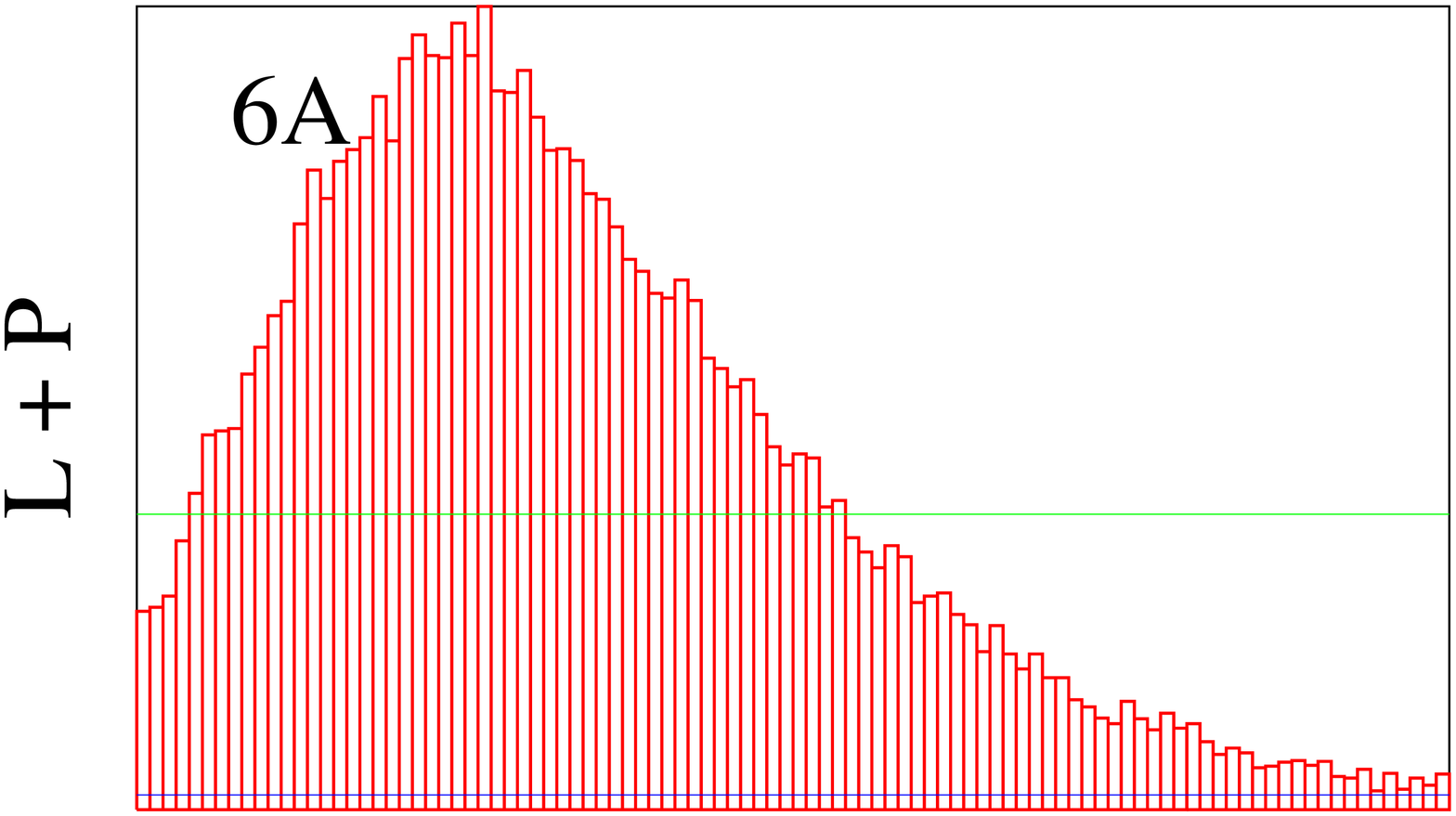}
\hspace{-6.5mm}
\includegraphics[%
scale=0.19,angle=270]{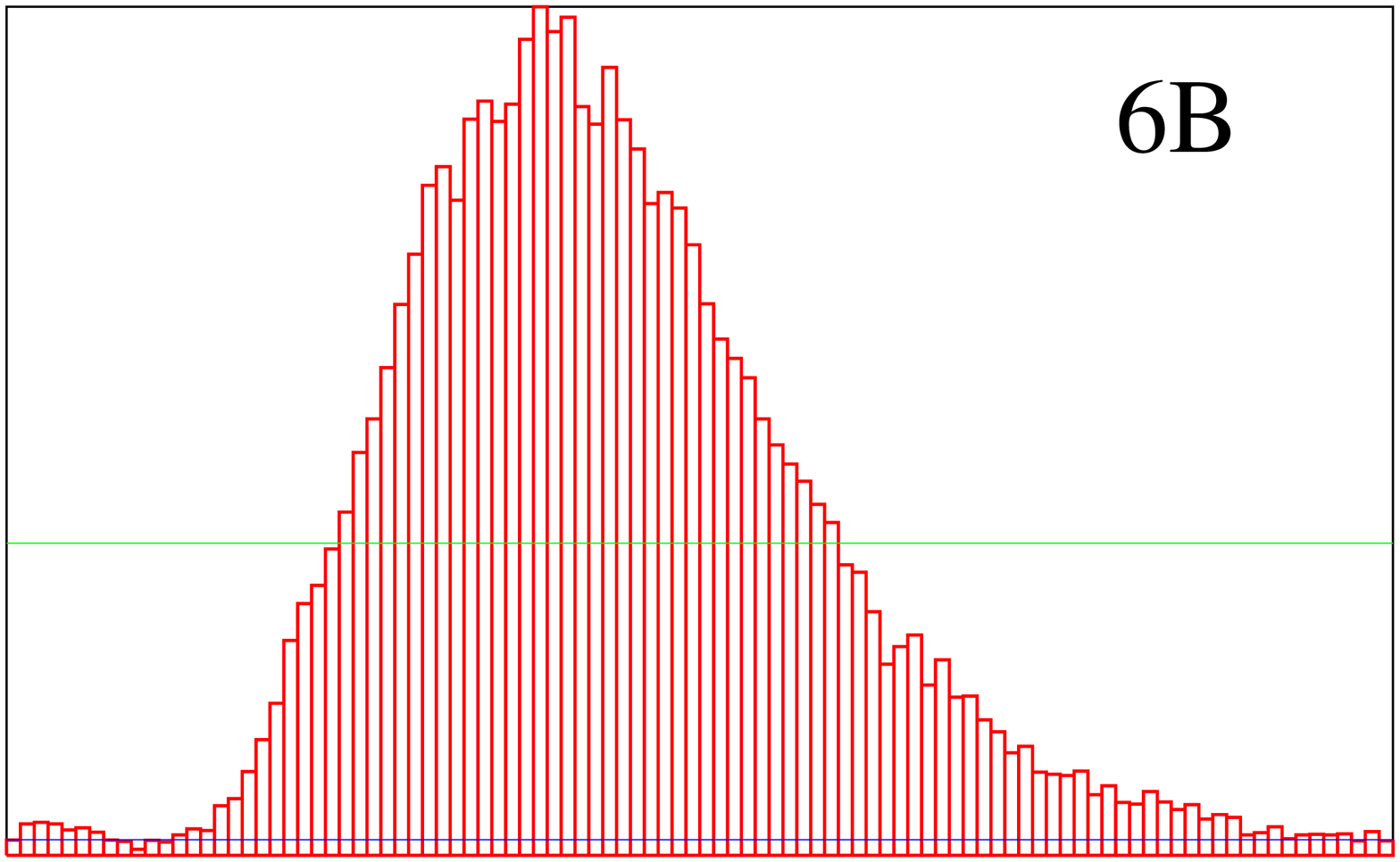}
\hspace{-6.5mm}
\includegraphics[%
scale=0.19,angle=270]{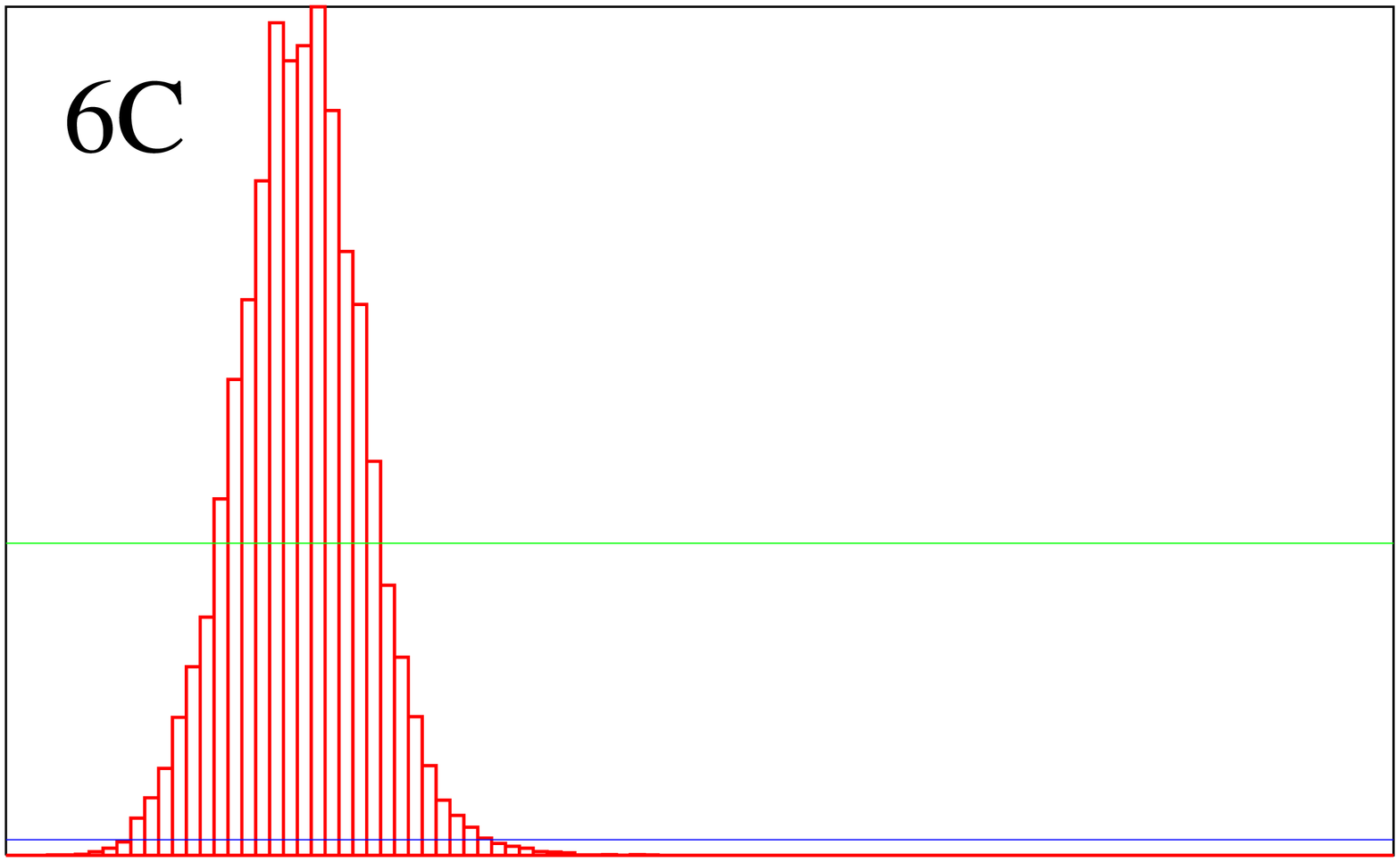}
\vspace{-9.8mm}
\tabularnewline
\includegraphics[%
scale=0.19,angle=270]{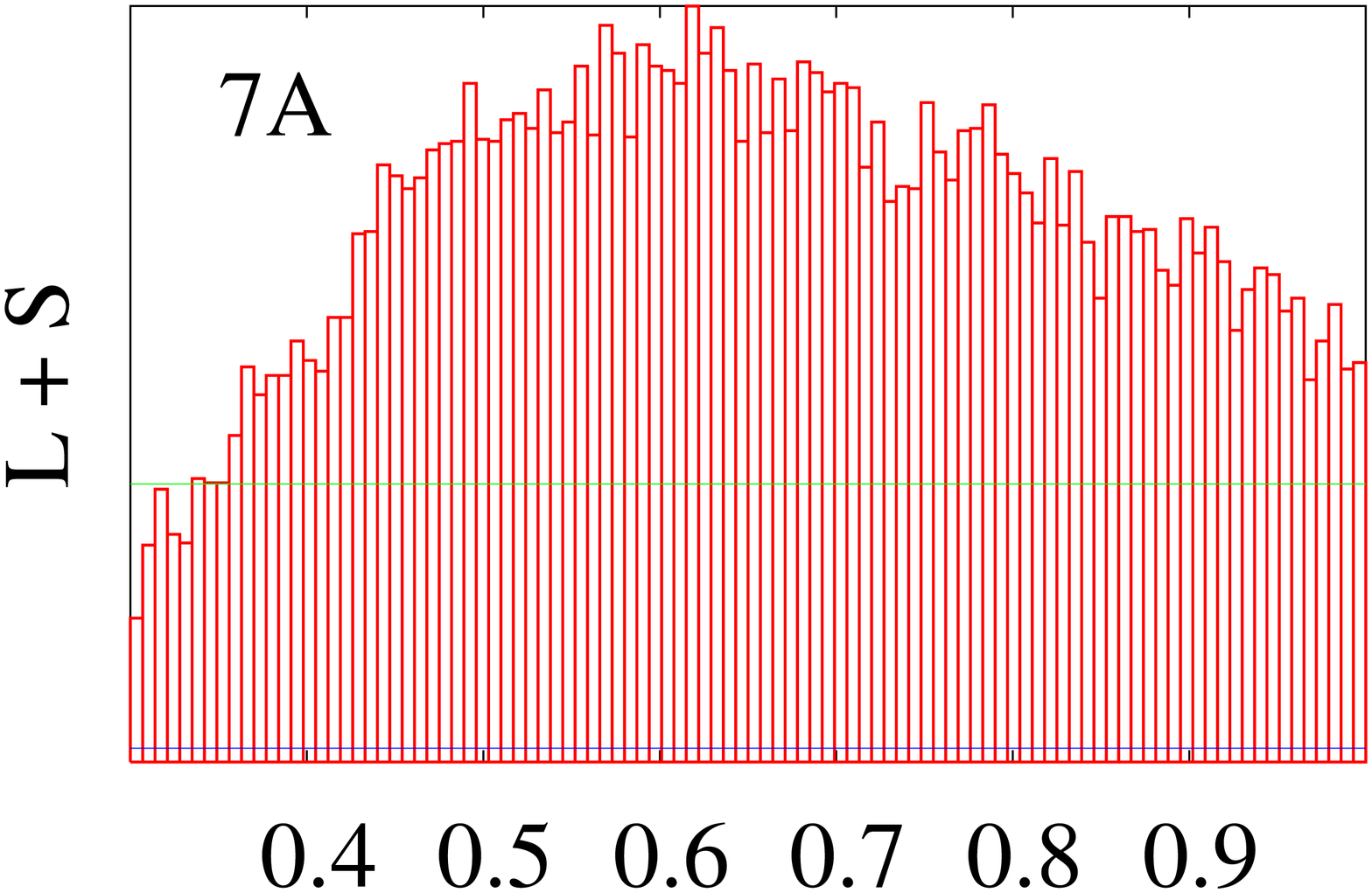}
\hspace{-6.5mm}
\includegraphics[%
scale=0.19,angle=270]{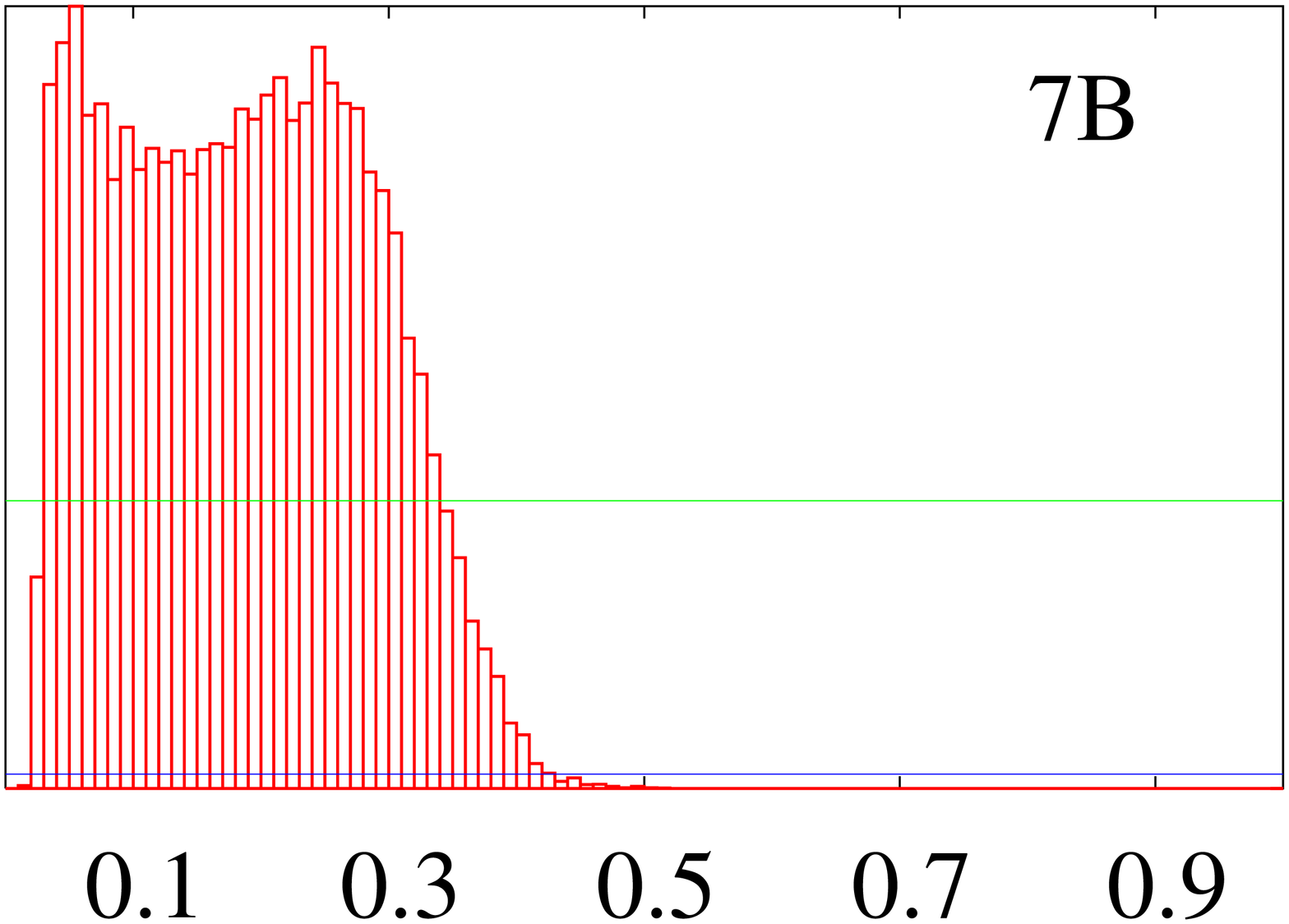}
\hspace{-6.5mm}
\includegraphics[%
scale=0.19,angle=270]{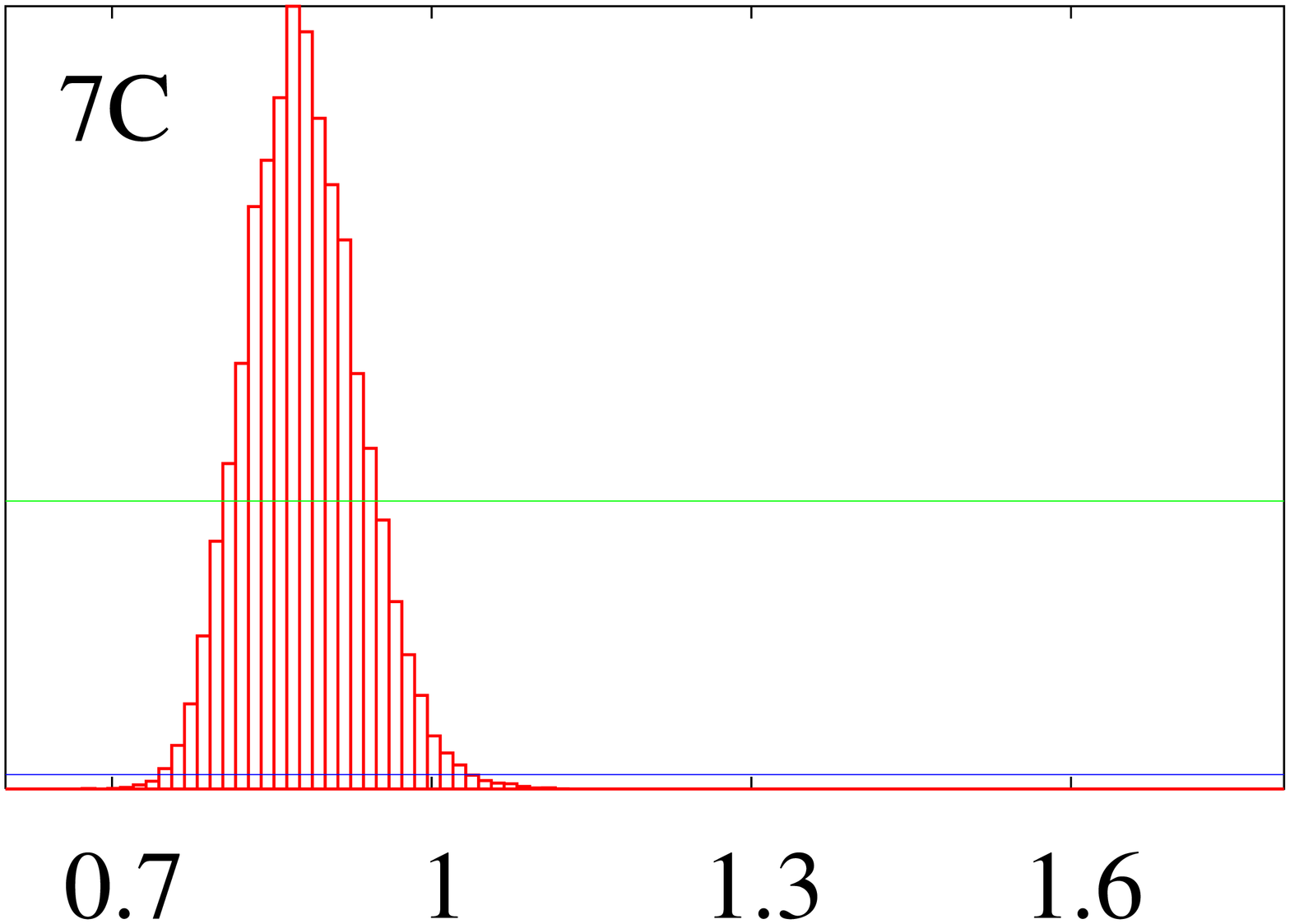}
\vspace{-2mm}
\tabularnewline
\end{tabular}
\end{center}
\caption{Constraints on the Hubble parameter $h$ (first column), the matter density $\Omega_\textrm{m}$ (second column) and the mass density fluctuation parameter $\sigma_8$ (third column) obtained from different combinations of large-scale structure data and supernovae. Each row corresponds to a different combination. The first four rows are each data set separately. The supernovae likelihood function does not constrain $\sigma_8$.}
\label{lss-histograms-1}
\end{figure}

\begin{figure}[!bht]
\begin{center}
\begin{tabular}{ccccccccc}
\tabularnewline 
\includegraphics[%
scale=0.19,angle=270]{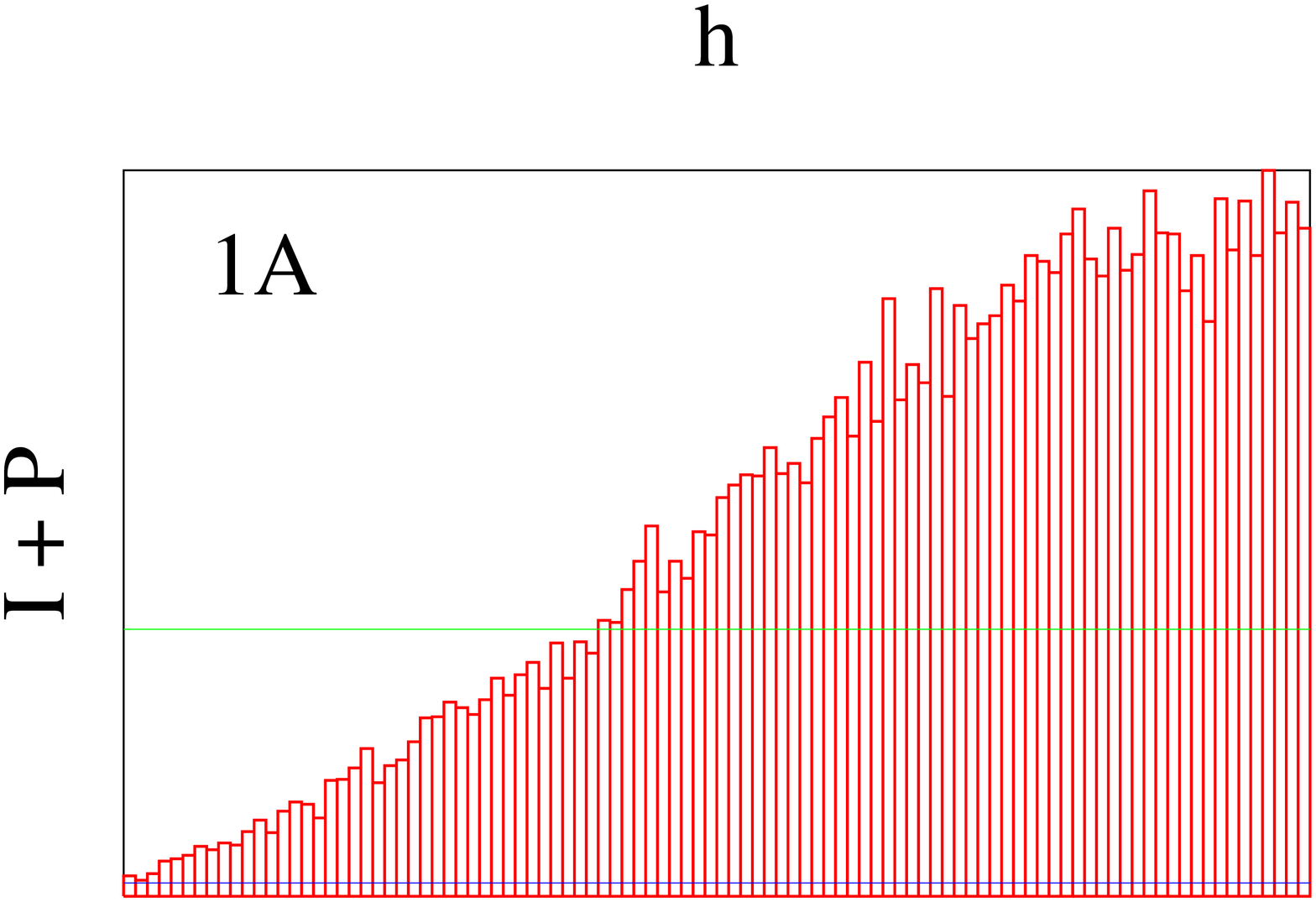}
\hspace{-6.5mm}
\includegraphics[%
scale=0.19,angle=270]{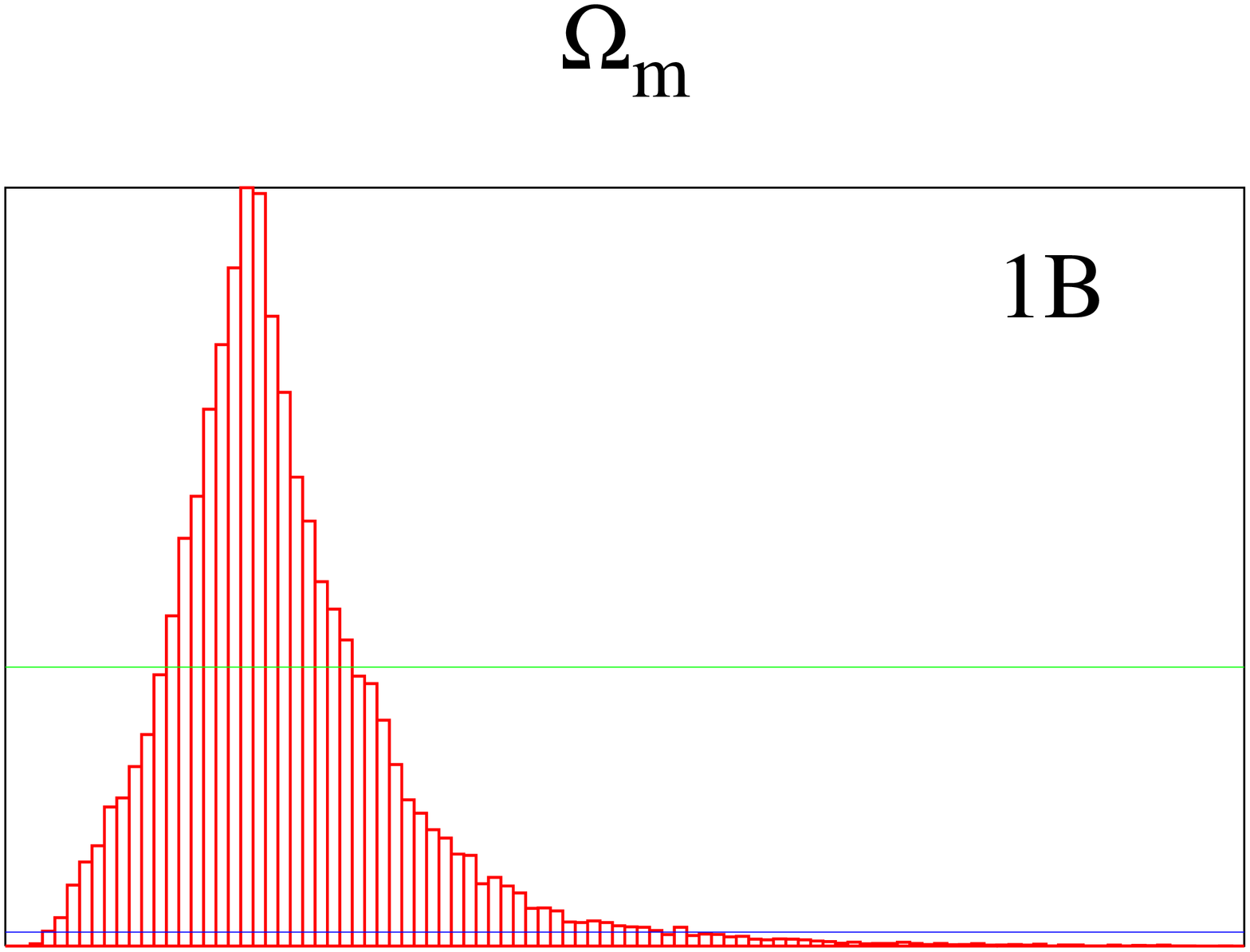}
\hspace{-6.5mm}
\includegraphics[%
scale=0.19,angle=270]{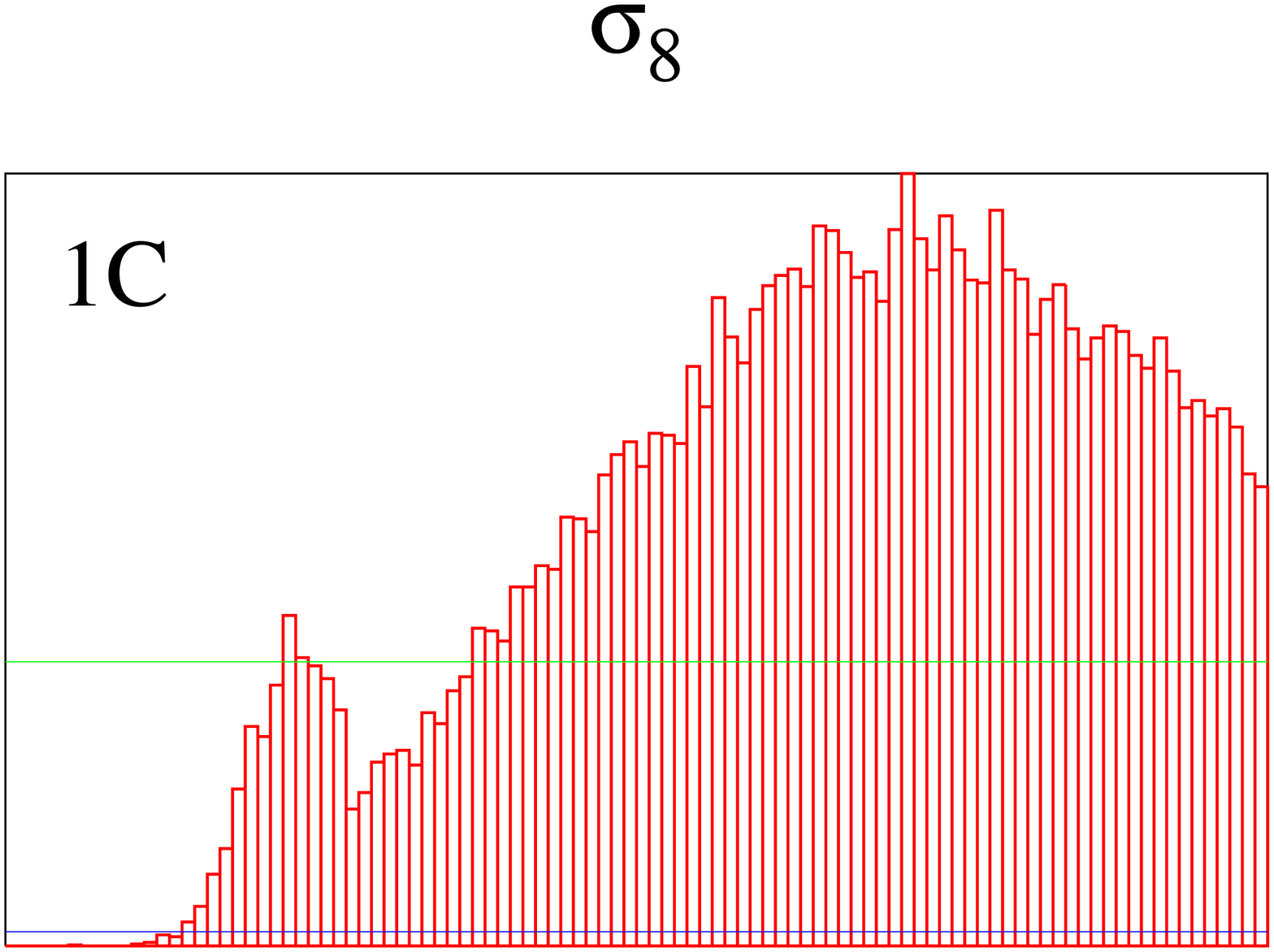}
\vspace{-9.8mm}
\tabularnewline
\includegraphics[%
scale=0.19,angle=270]{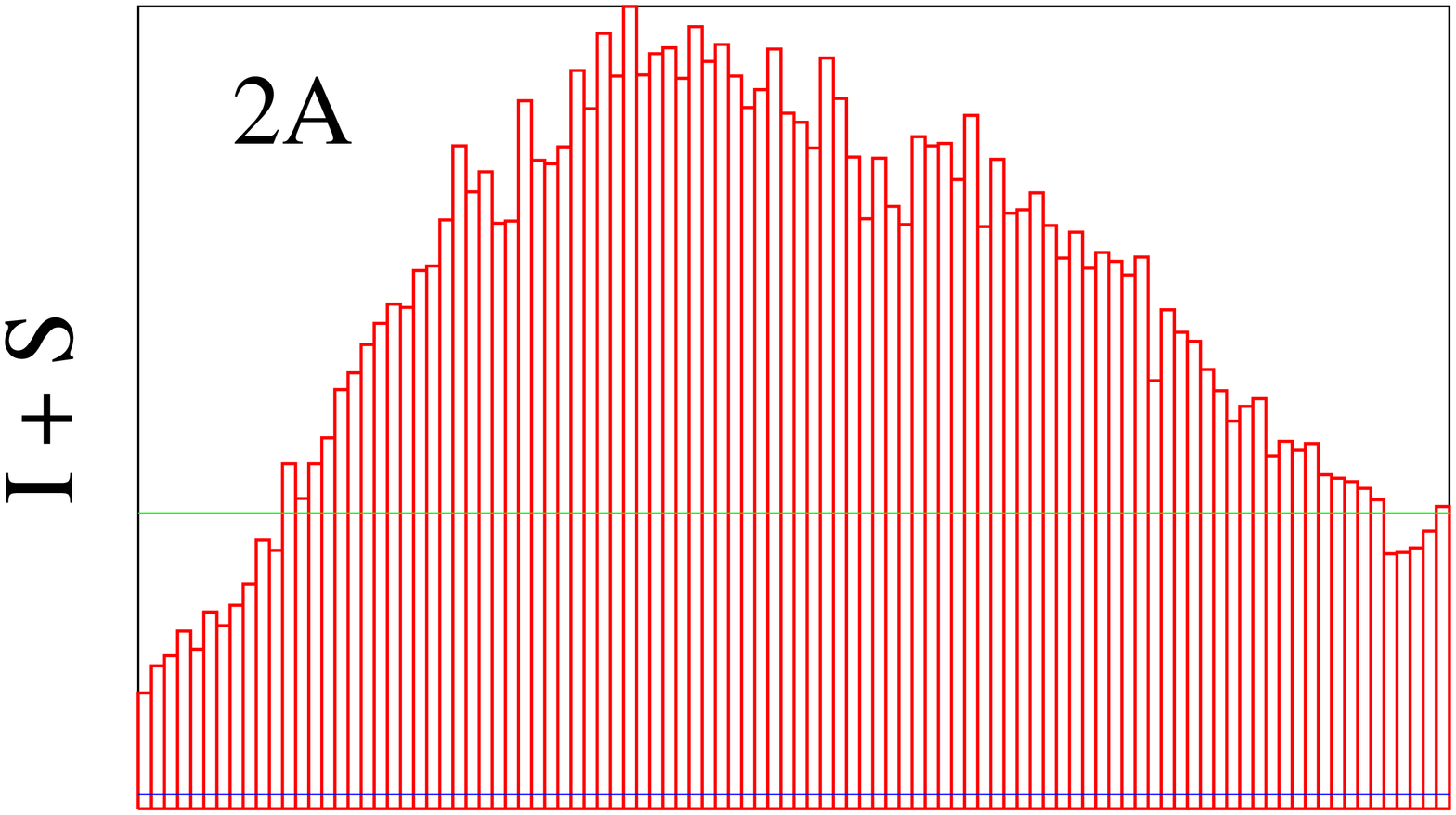}
\hspace{-6.5mm}
\includegraphics[%
scale=0.19,angle=270]{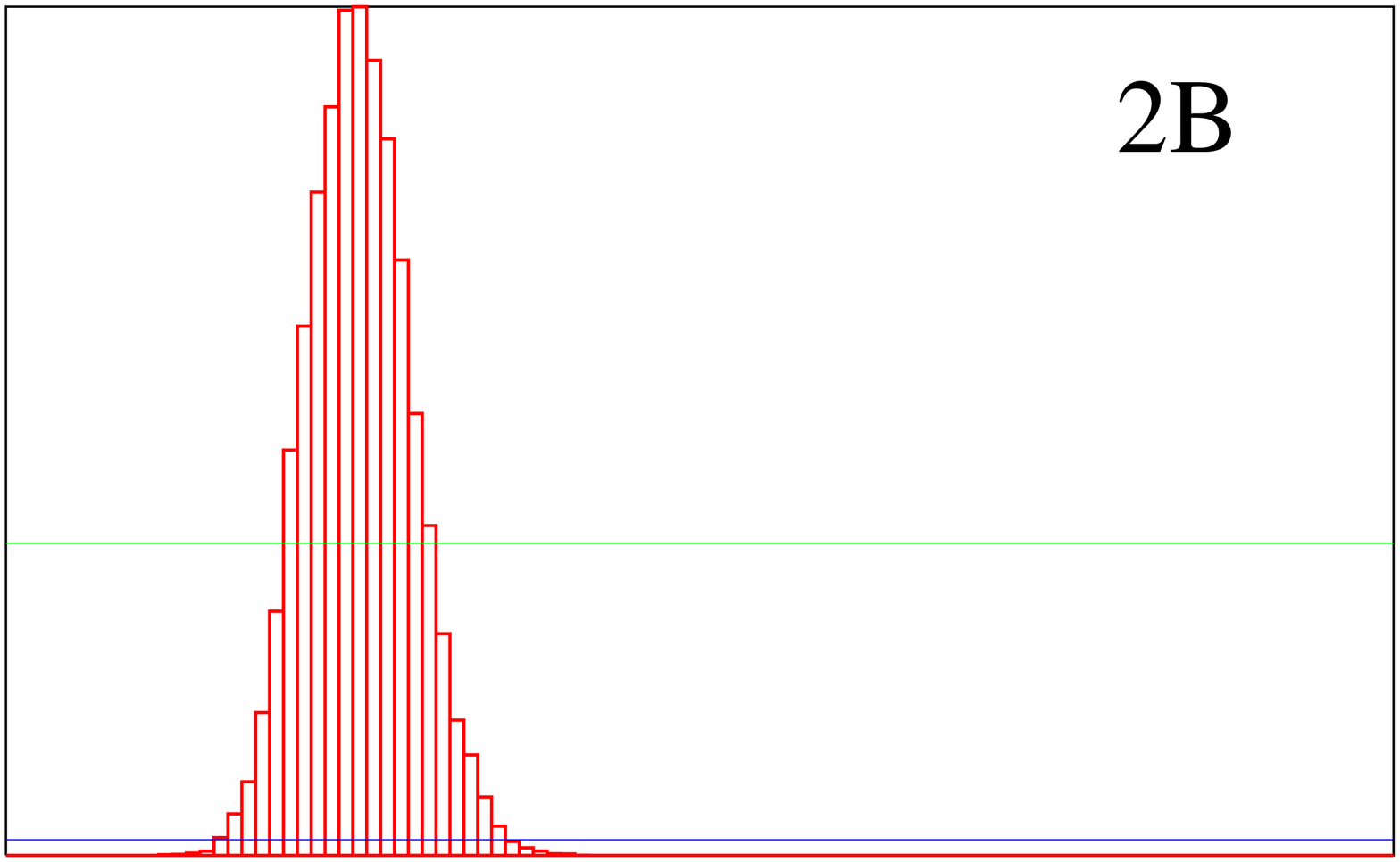}
\hspace{-6.5mm}
\includegraphics[%
scale=0.19,angle=270]{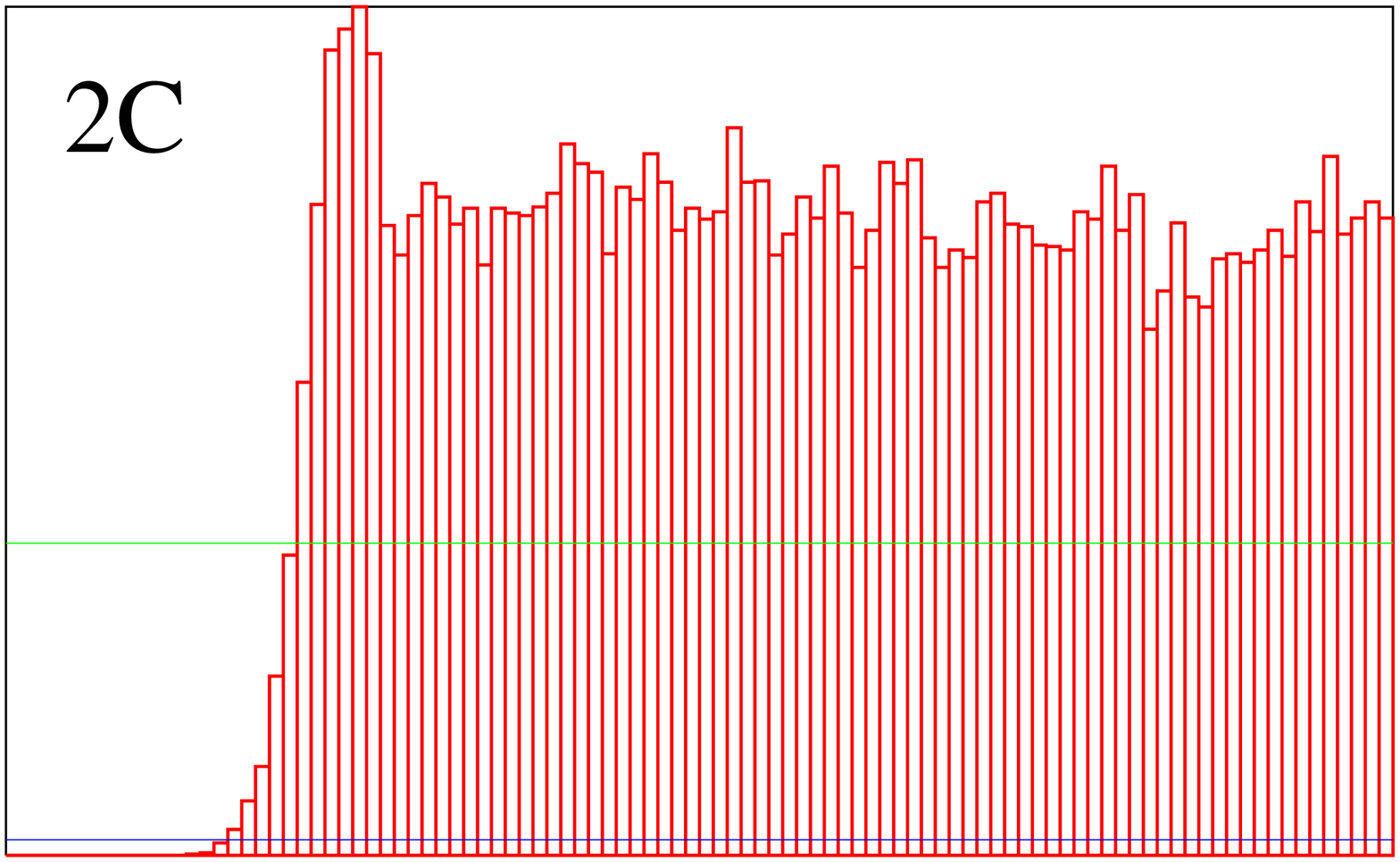}
\vspace{-9.8mm}
\tabularnewline
\includegraphics[%
scale=0.19,angle=270]{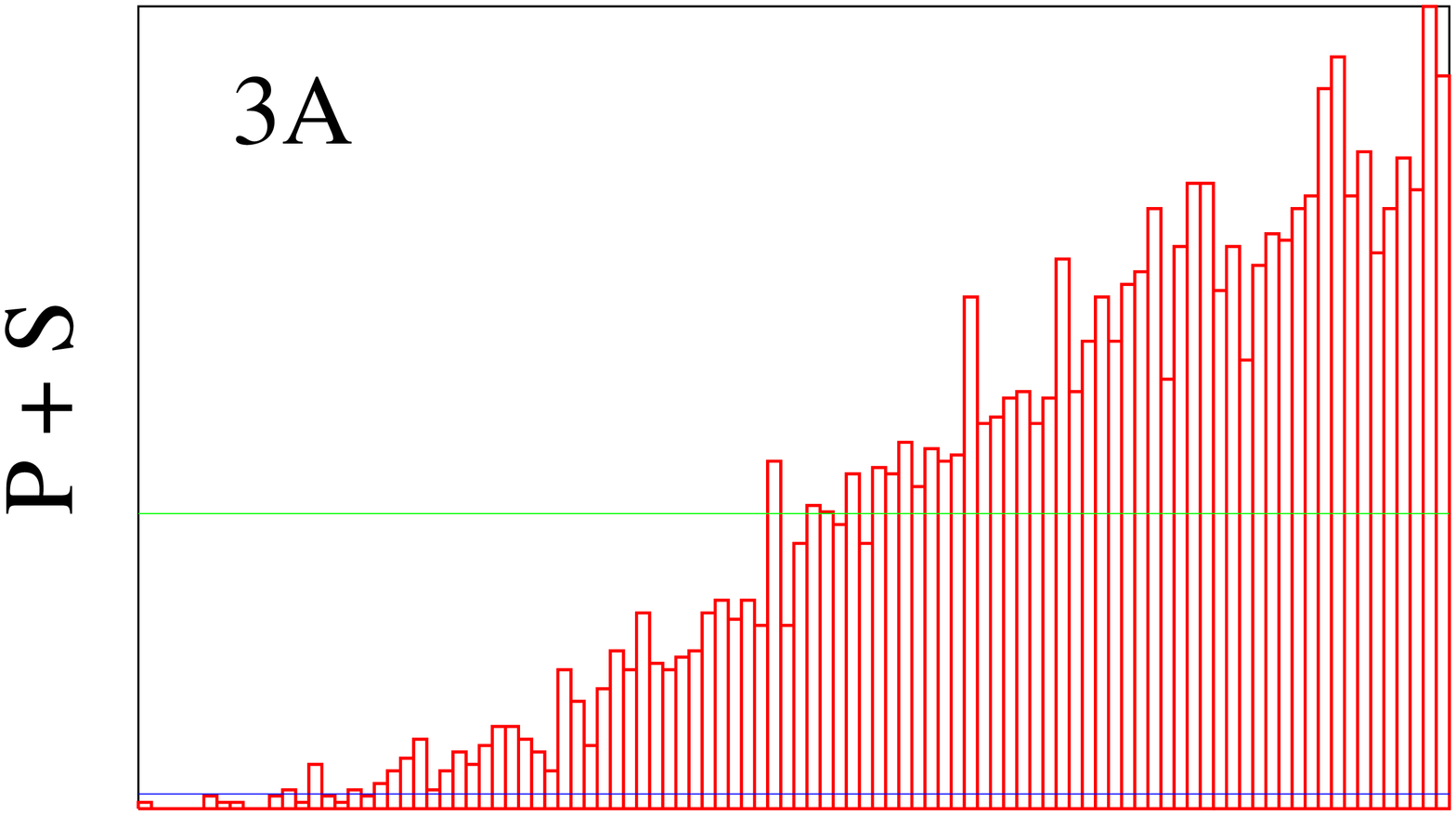}
\hspace{-6.5mm}
\includegraphics[%
scale=0.19,angle=270]{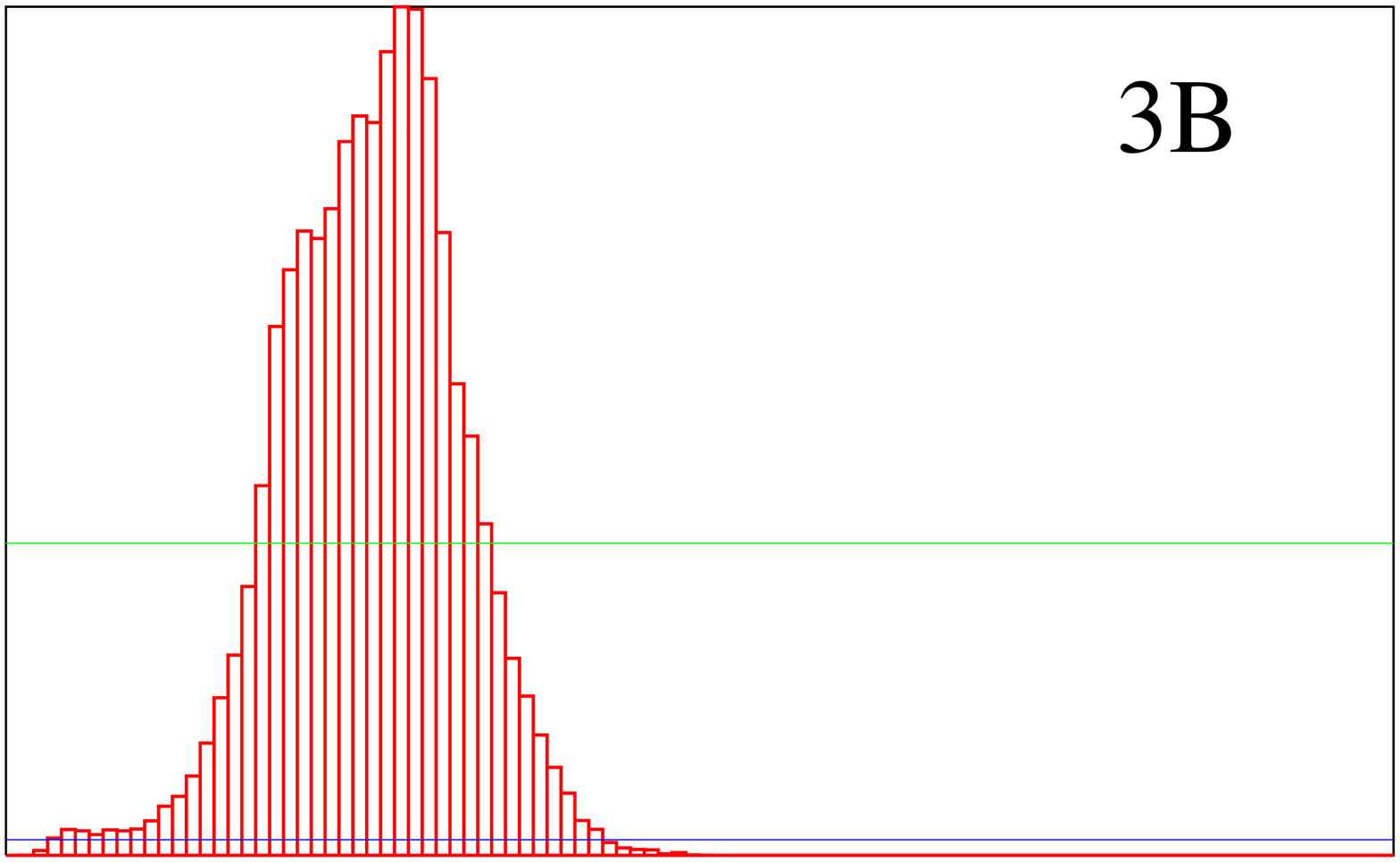}
\hspace{-6.5mm}
\includegraphics[%
scale=0.19,angle=270]{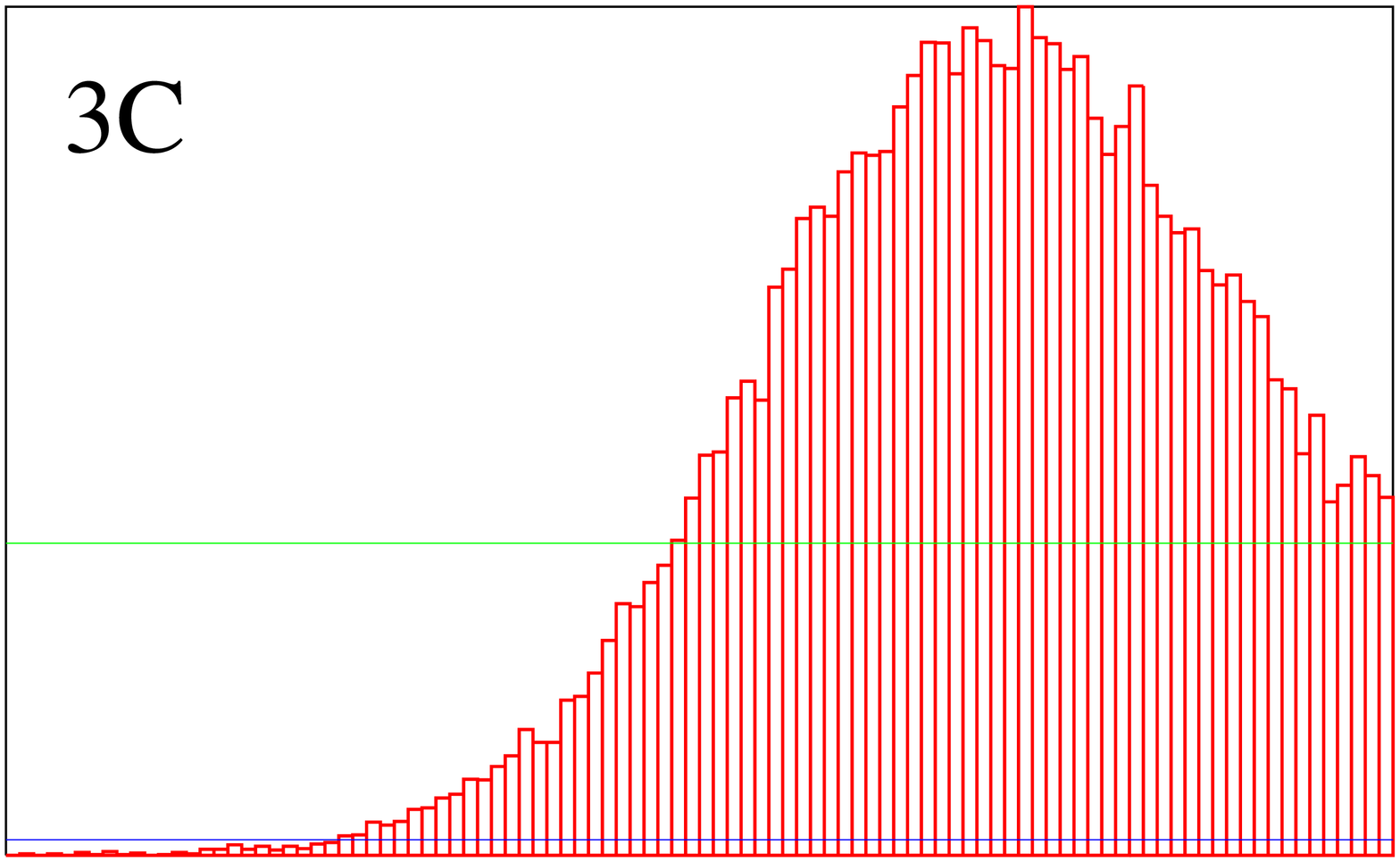}
\vspace{-9.8mm}
\tabularnewline
\includegraphics[%
scale=0.19,angle=270]{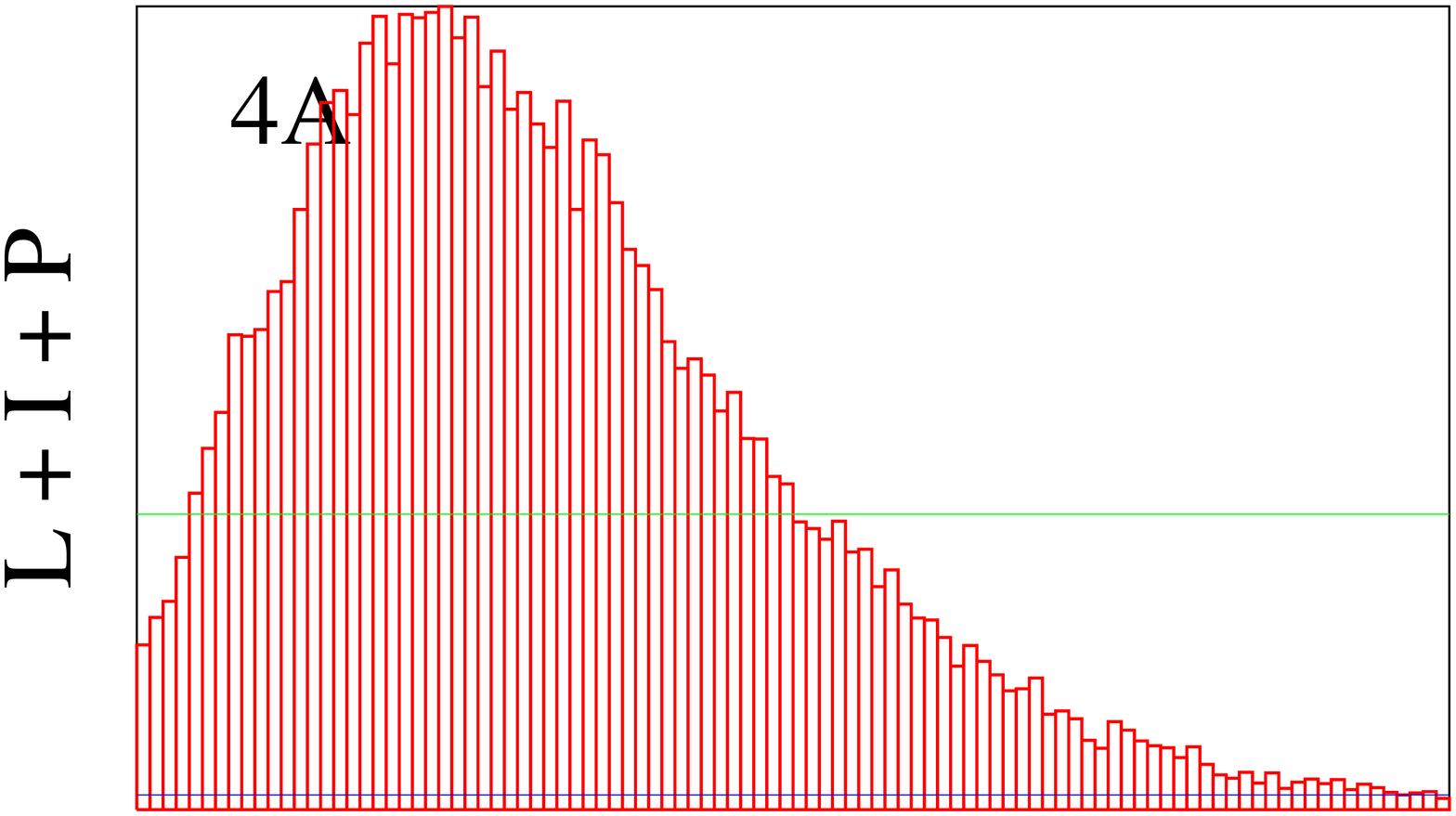}
\hspace{-6.5mm}
\includegraphics[%
scale=0.19,angle=270]{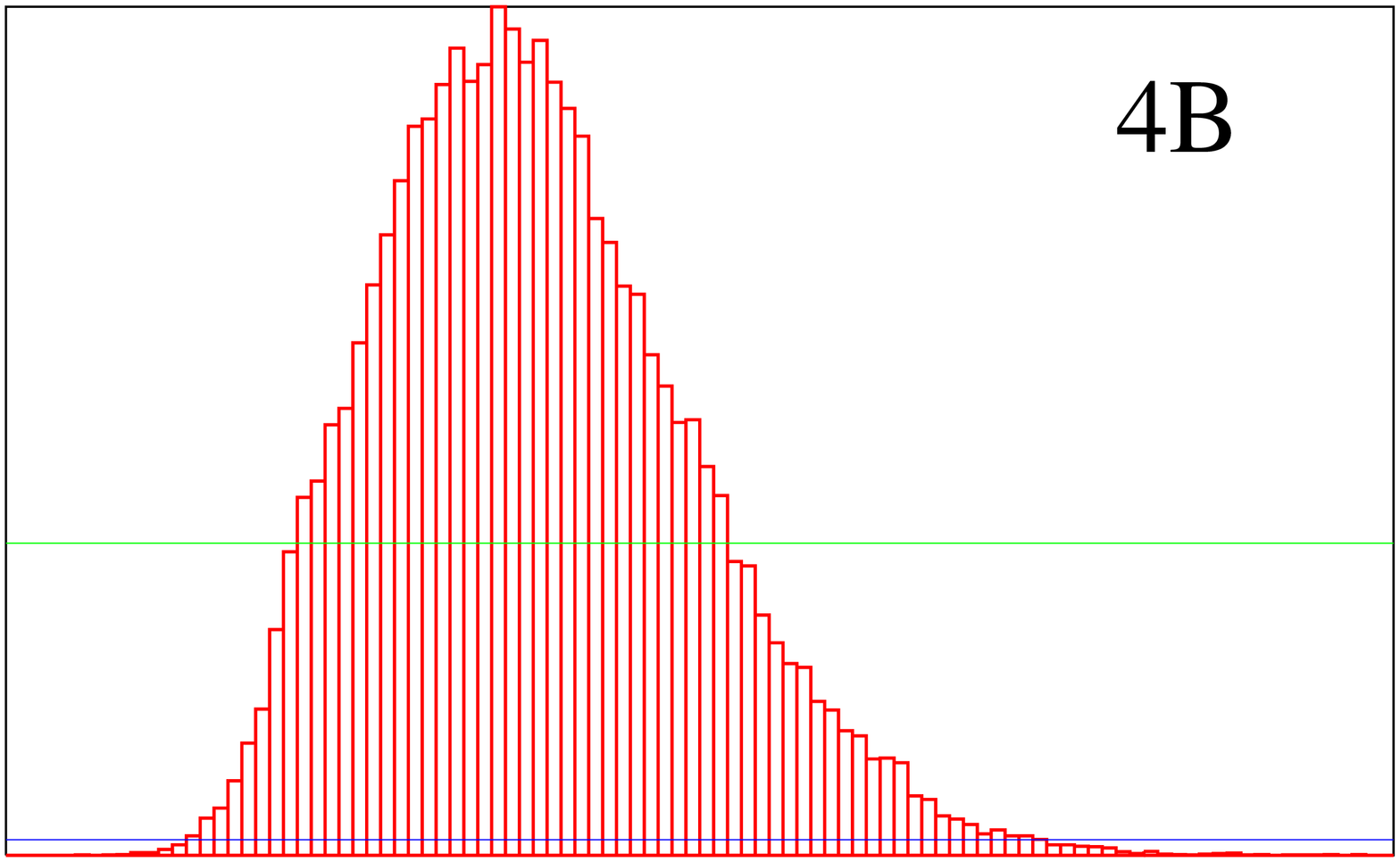}
\hspace{-6.5mm}
\includegraphics[%
scale=0.19,angle=270]{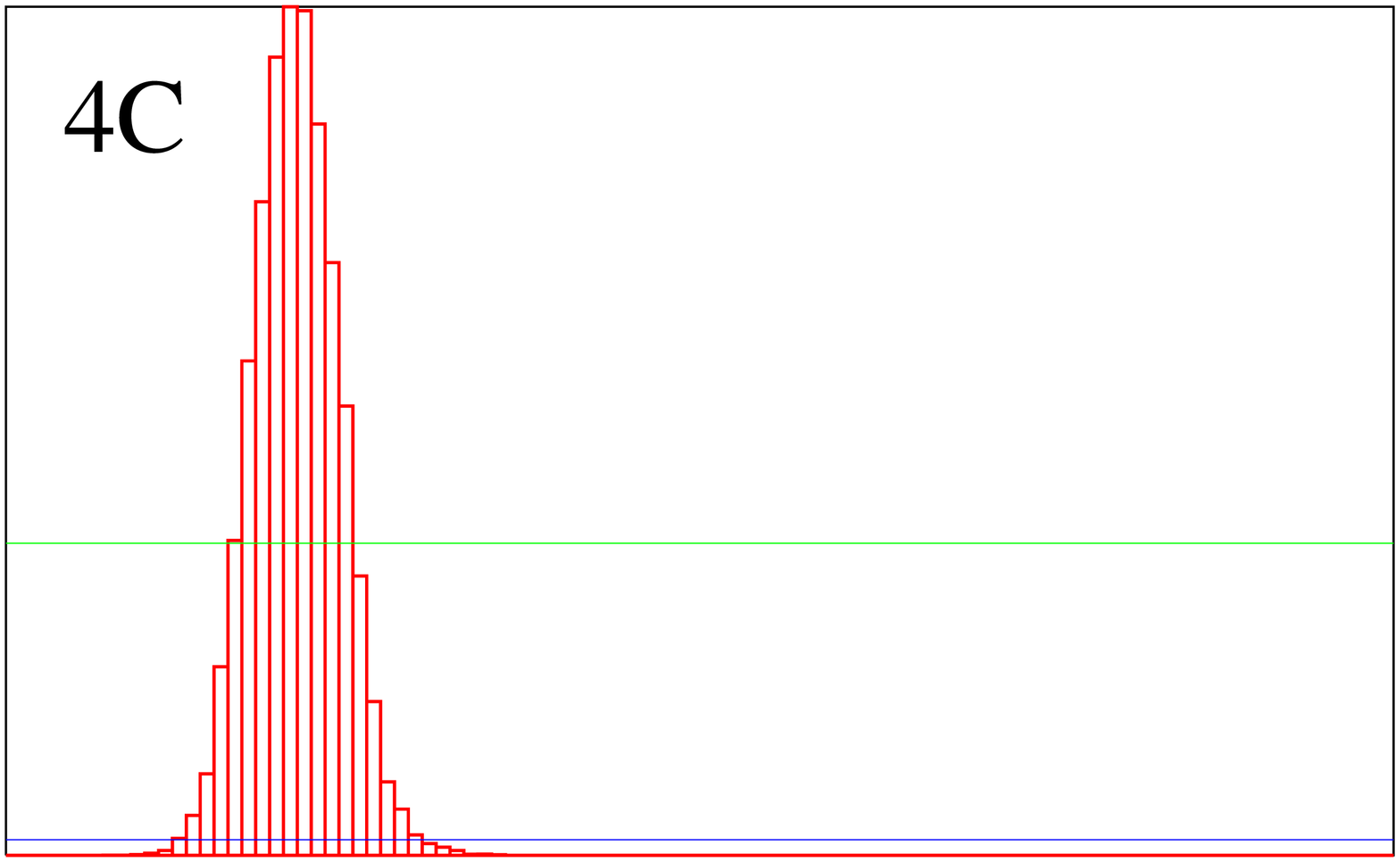}
\vspace{-9.8mm}
\tabularnewline
\includegraphics[%
scale=0.19,angle=270]{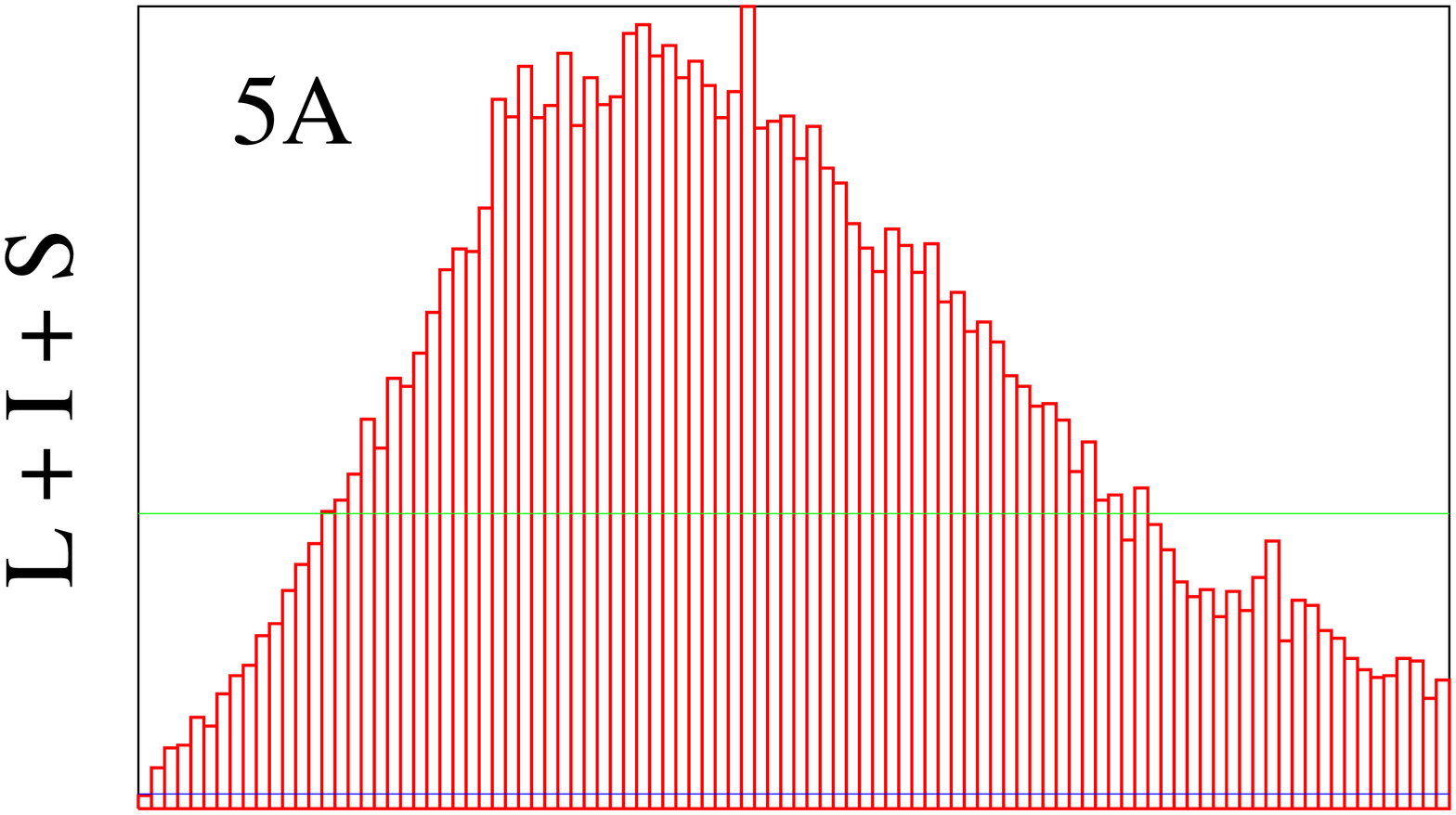}
\hspace{-6.5mm}
\includegraphics[%
scale=0.19,angle=270]{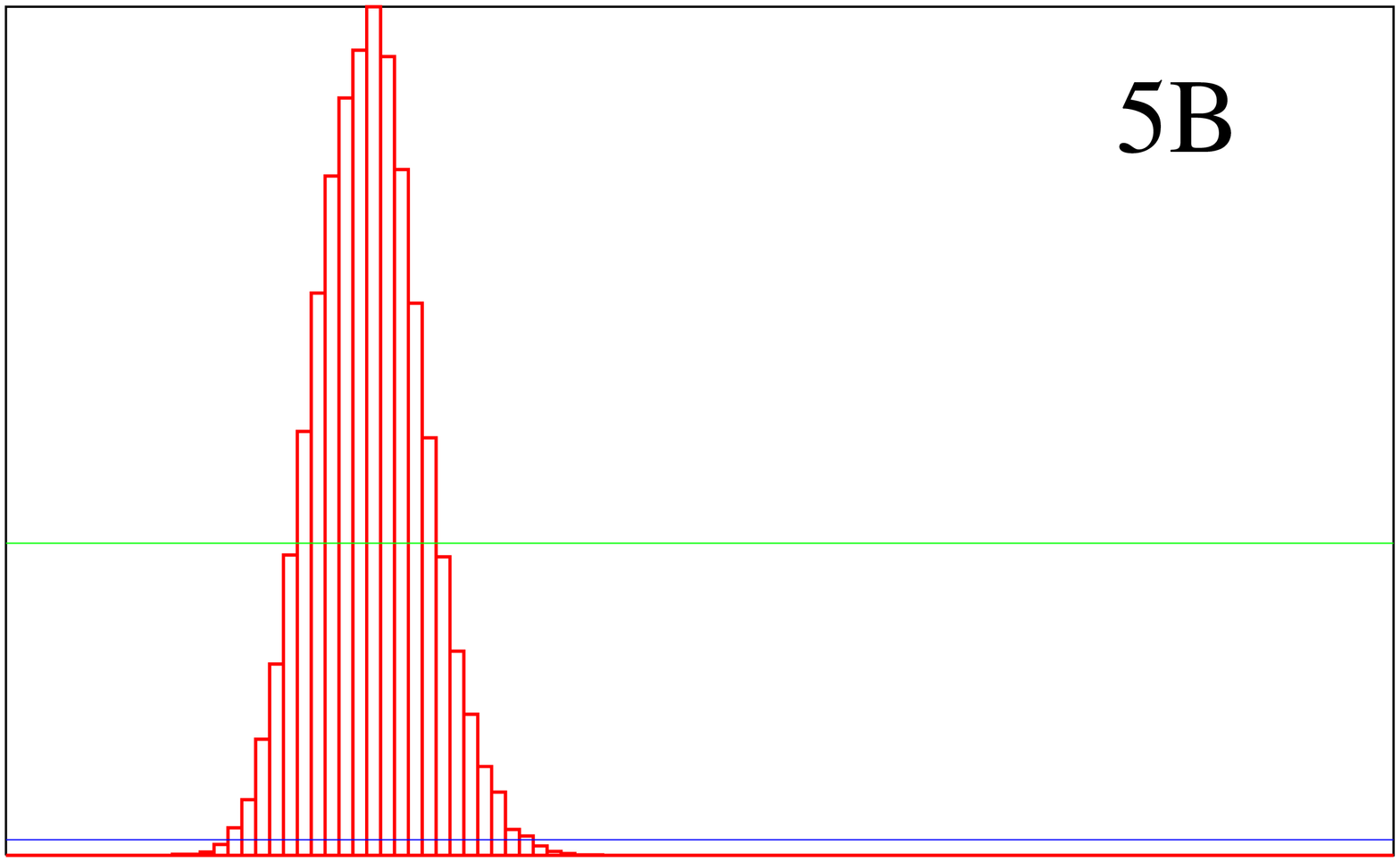}
\hspace{-6.5mm}
\includegraphics[%
scale=0.19,angle=270]{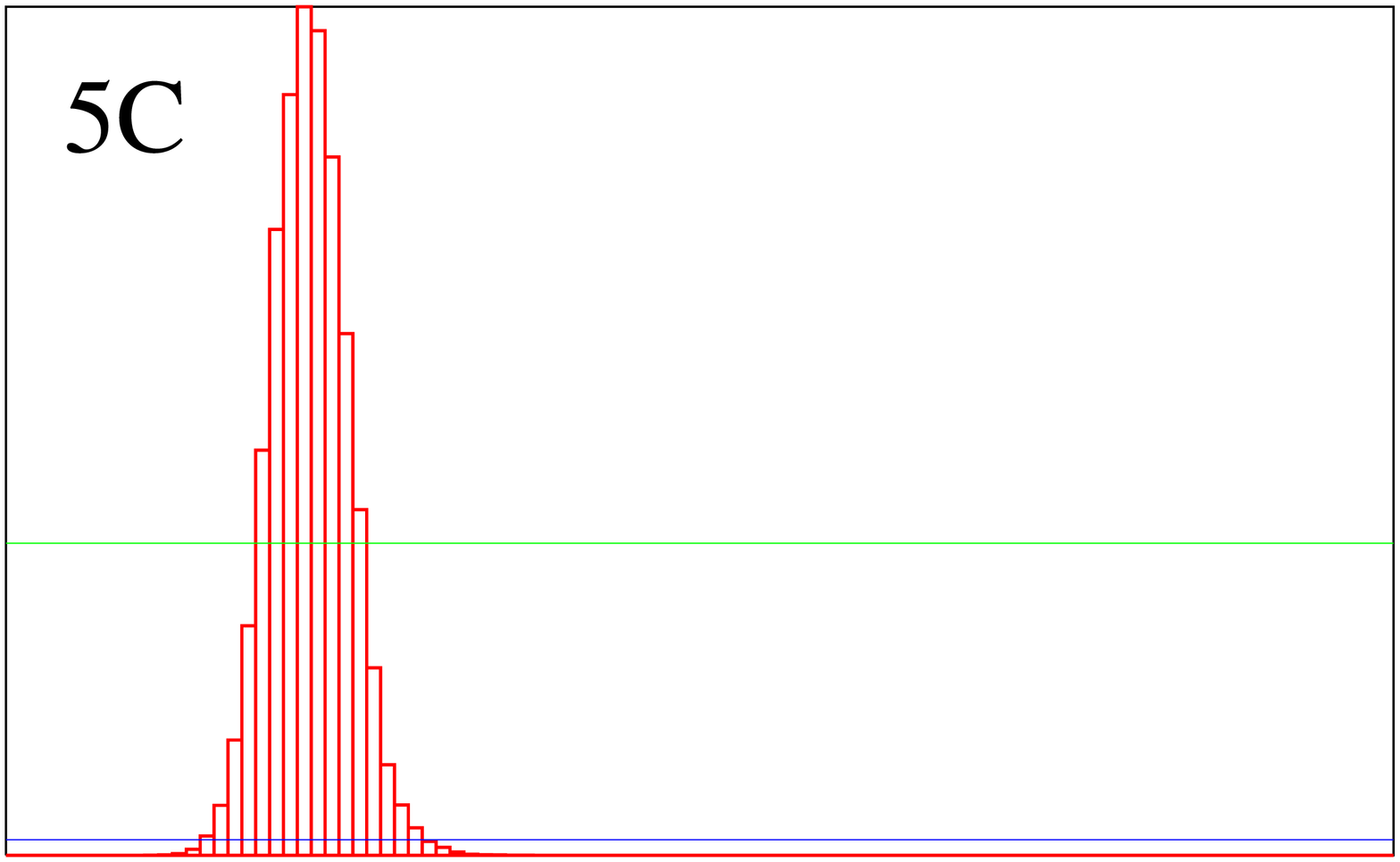}
\vspace{-9.8mm}
\tabularnewline
\includegraphics[%
scale=0.19,angle=270]{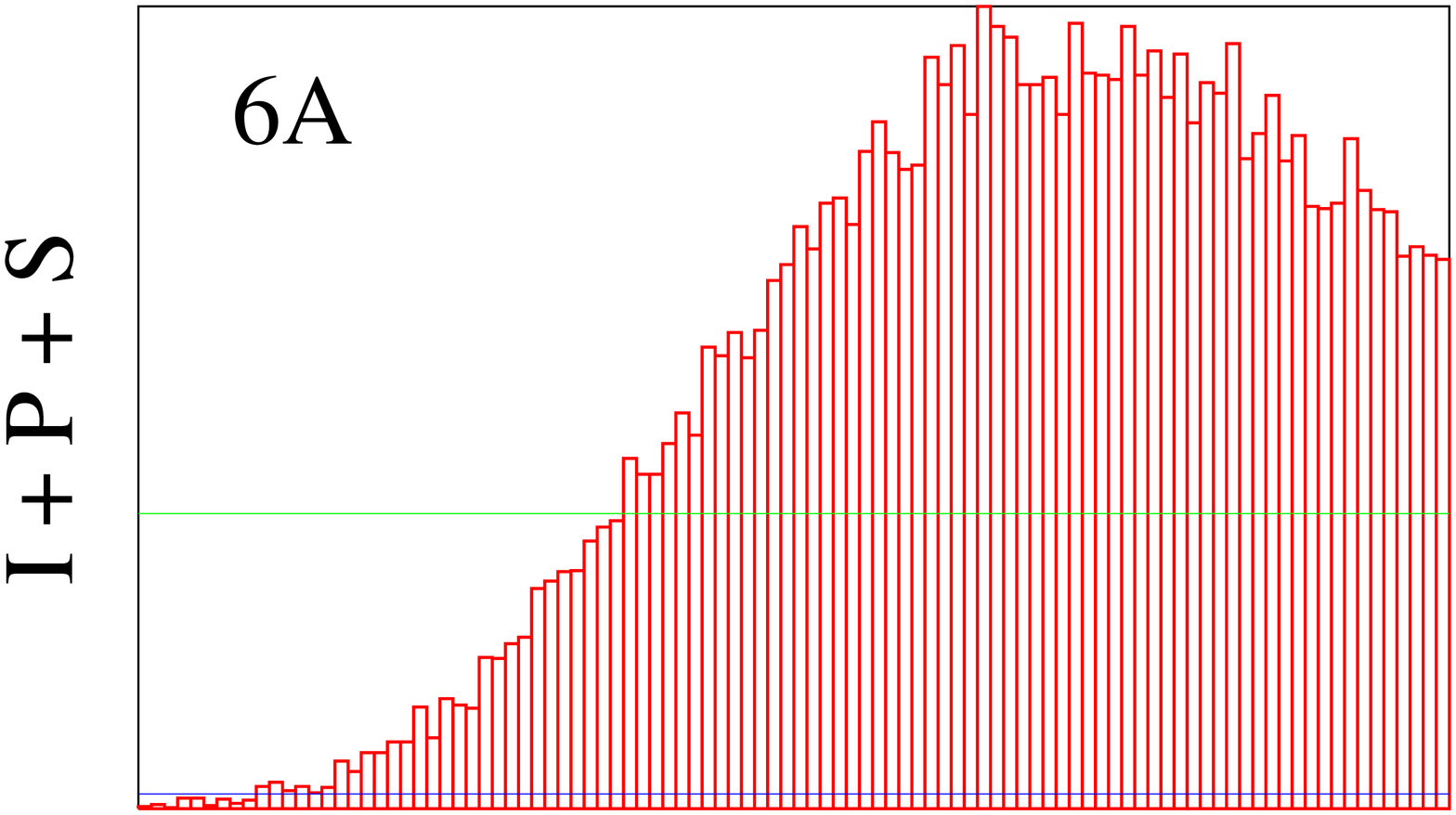}
\hspace{-6.5mm}
\includegraphics[%
scale=0.19,angle=270]{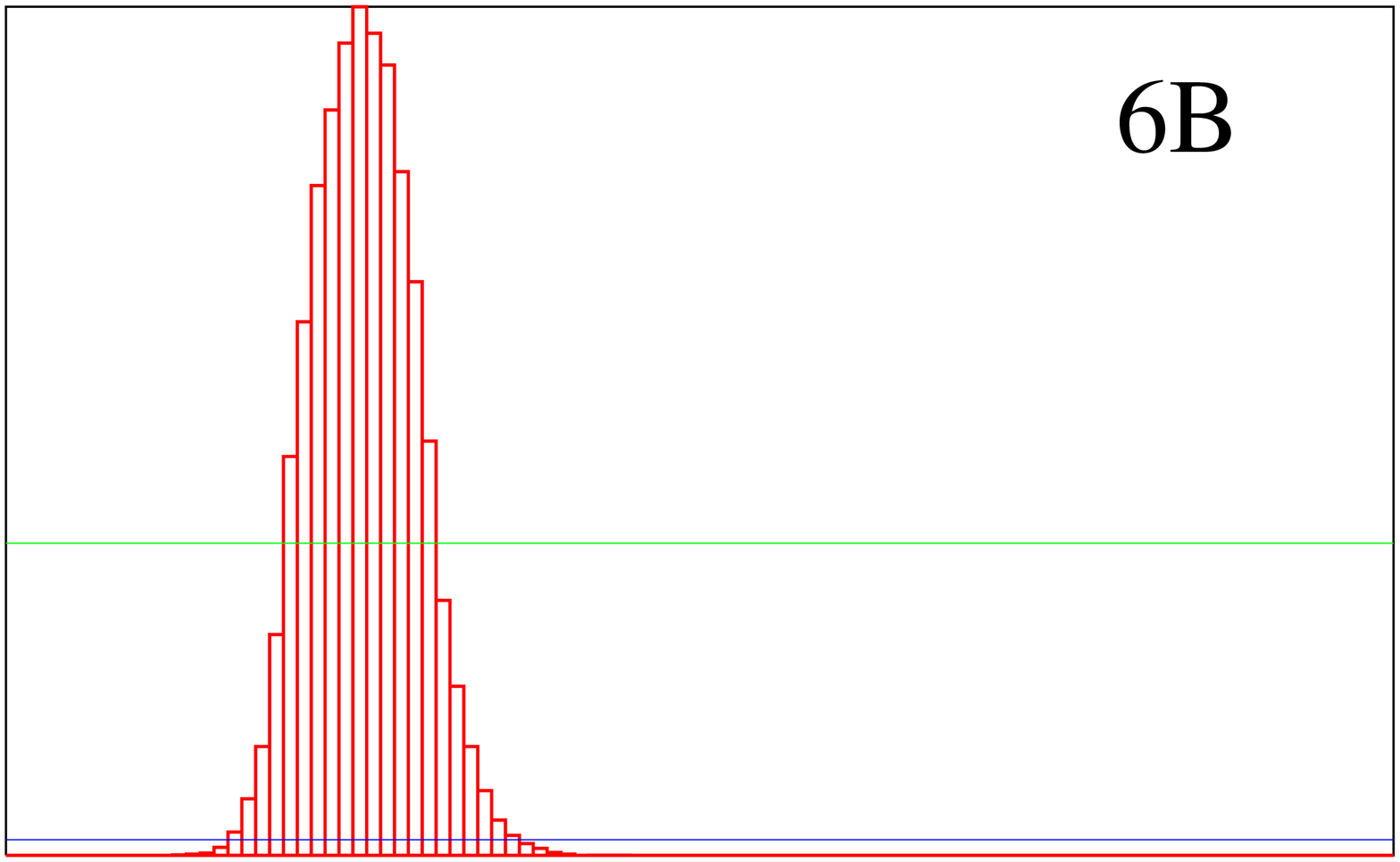}
\hspace{-6.5mm}
\includegraphics[%
scale=0.19,angle=270]{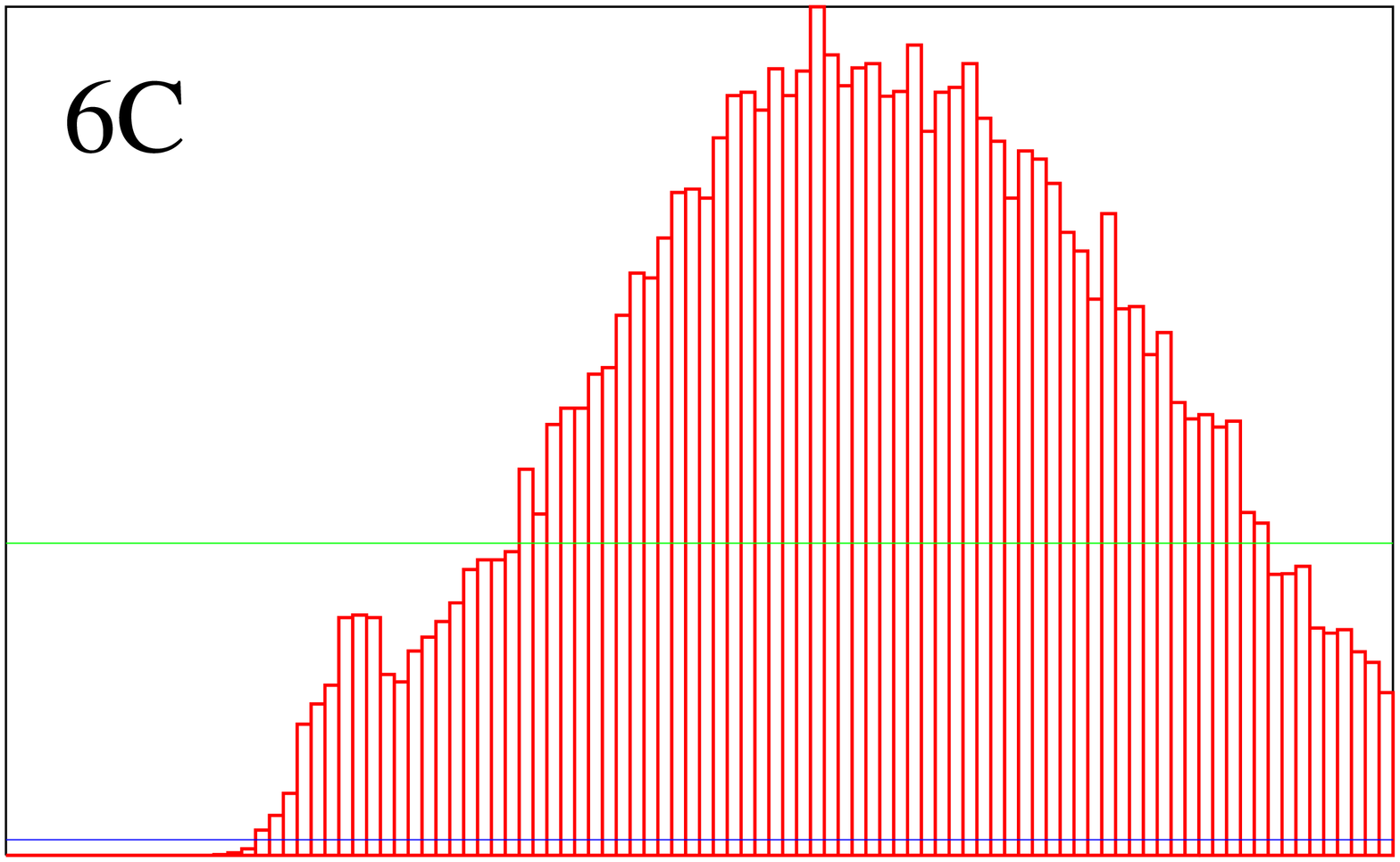}
\vspace{-9.8mm}
\tabularnewline
\includegraphics[%
scale=0.19,angle=270]{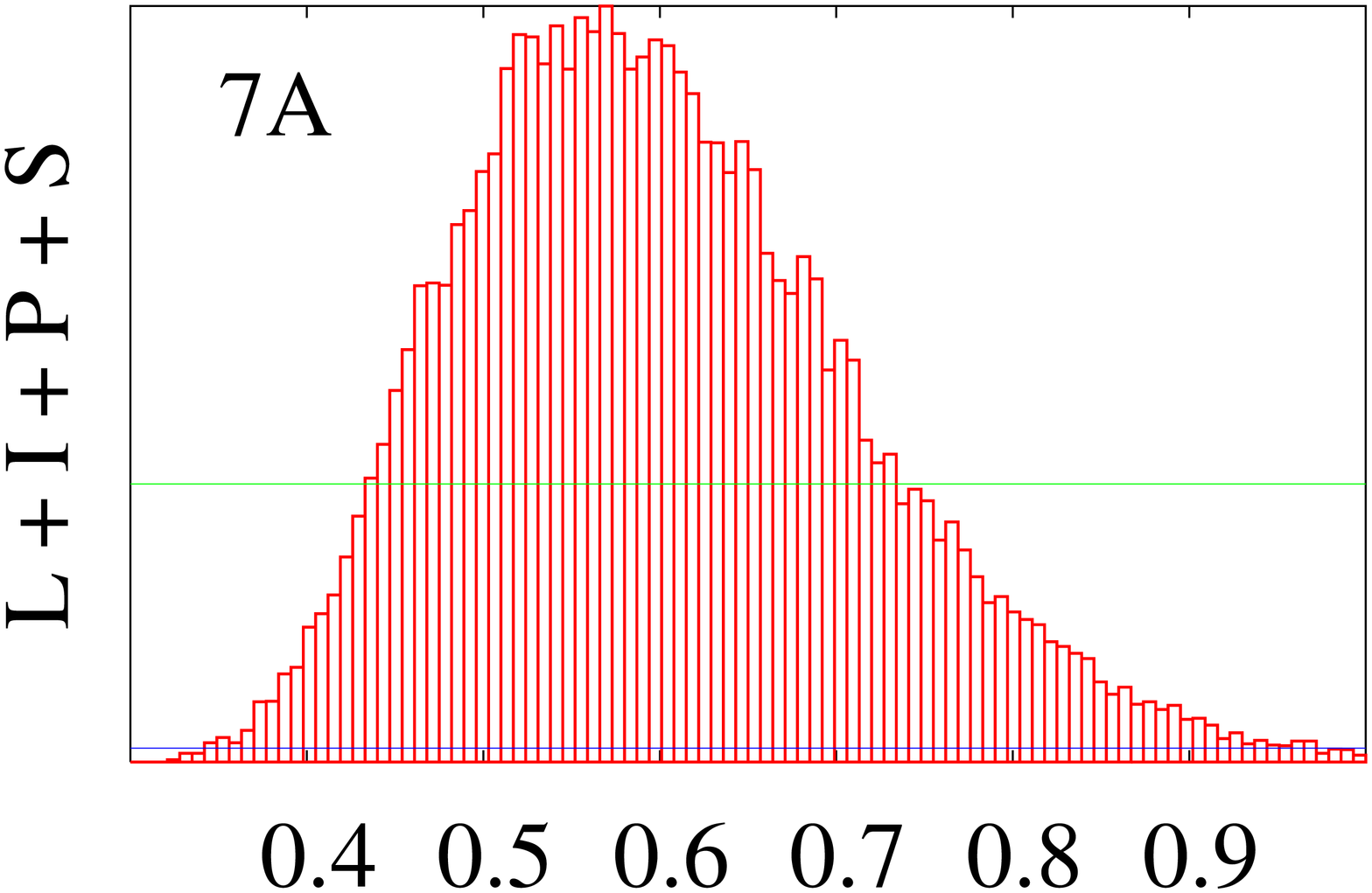}
\hspace{-6.5mm}
\includegraphics[%
scale=0.19,angle=270]{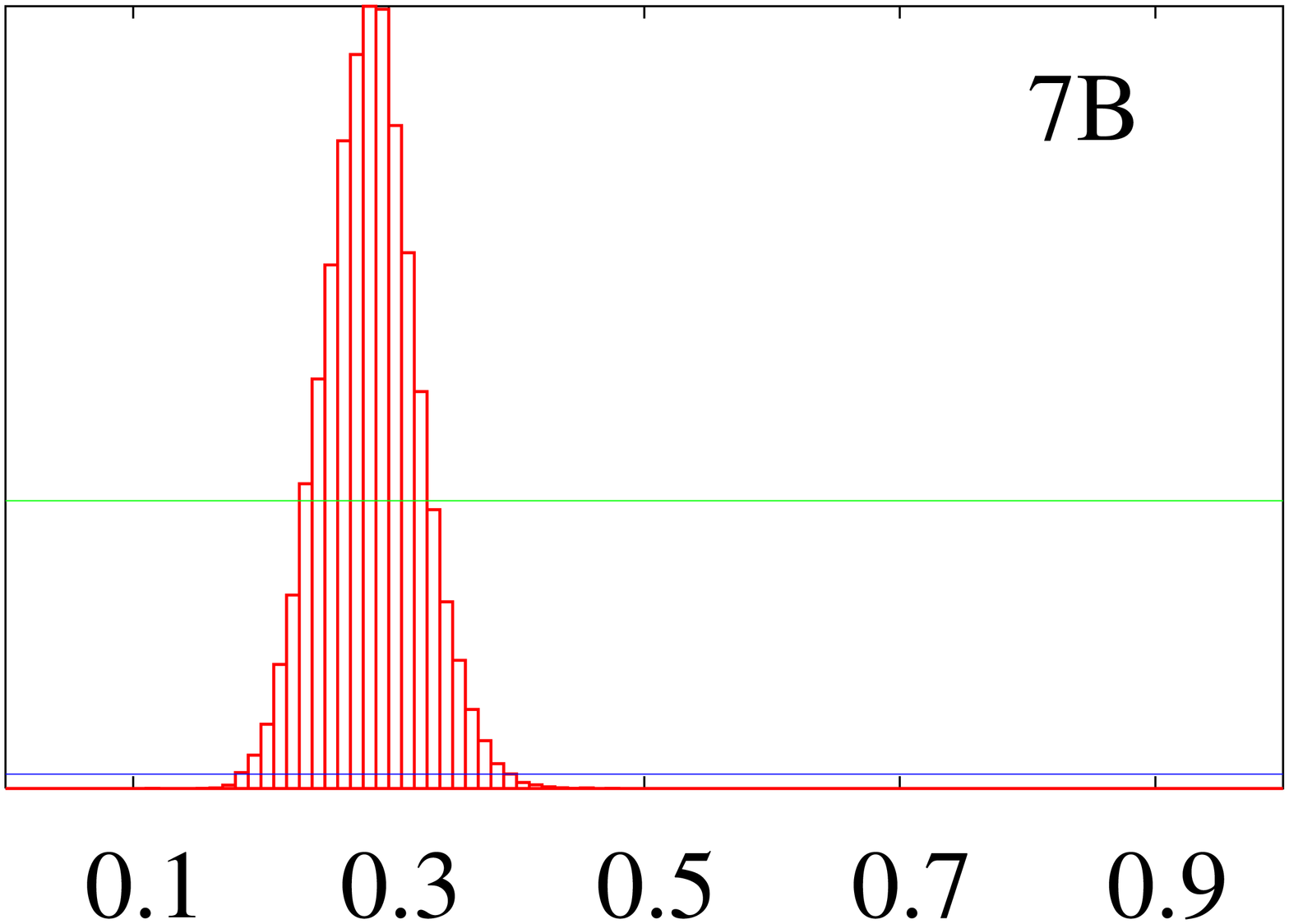}
\hspace{-6.5mm}
\includegraphics[%
scale=0.19,angle=270]{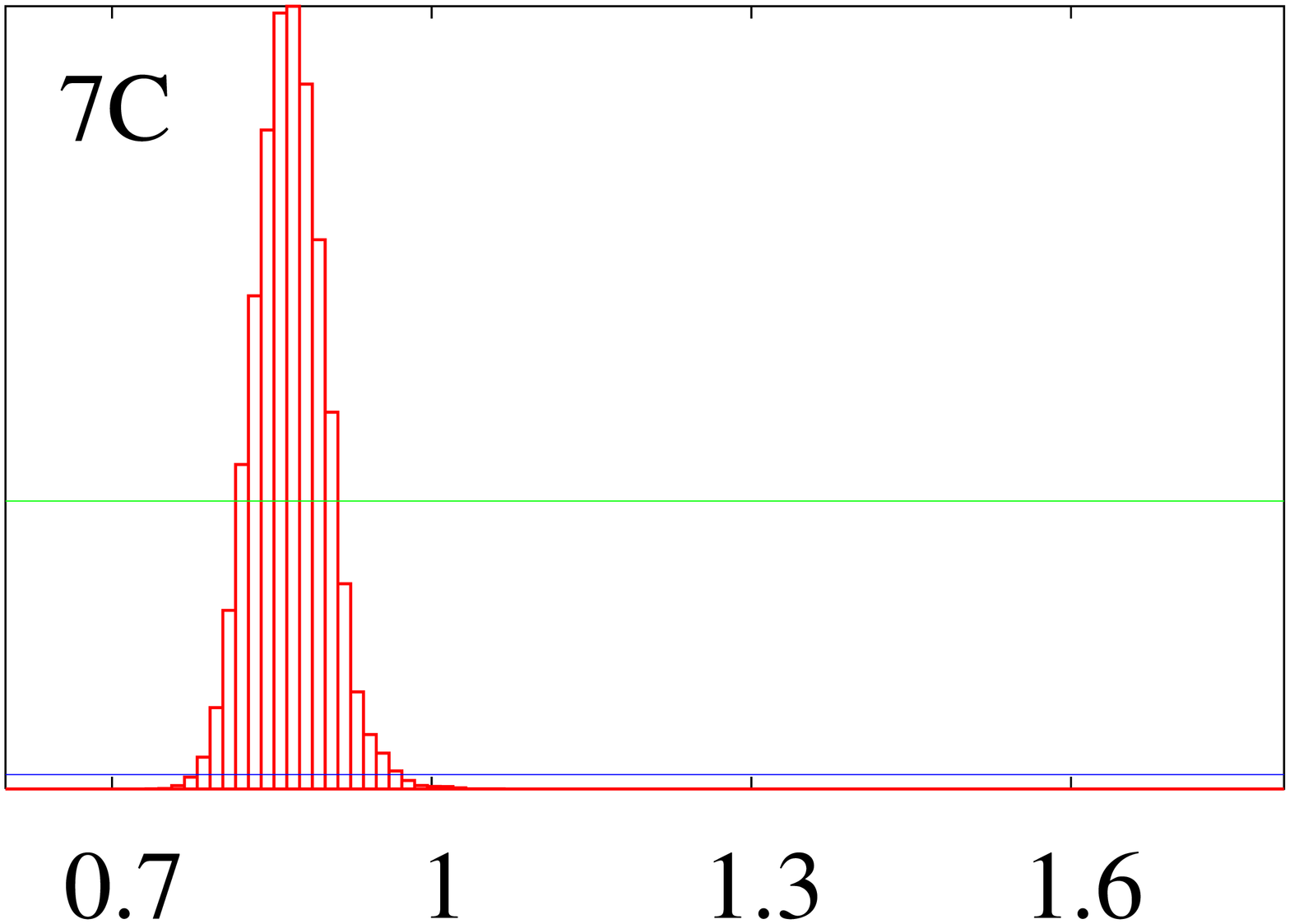}
\vspace{-2mm}
\tabularnewline
\end{tabular}
\end{center}
\caption{Constraints on the Hubble parameter $h$ (first column), the matter density $\Omega_\textrm{m}$ (second column) and the mass density fluctuation parameter $\sigma_8$ (third column) obtained from different combinations of large-scale structure data and supernovae.}
\label{lss-histograms-2}
\end{figure}

Table \ref{lss-results-table} lists the best fit values for $h$, $\Omega_\textrm{m}$ and $\sigma_8$ derived from the MCMC samples.  When available, the $1\sigma$ limits are given. ``--'' means the value was not constrained within the priors. The preferred values for the three parameters when all data sets are combined are $h=0.57_{-0.14}^{+0.16}$, $\Omega_\textrm{m} = 0.28 \pm 0.05$ and $\sigma_8 = 0.87_{-0.05}^{+0.04}$. This is in the case of a flat universe so $\Omega_{\Lambda} = 0.72 \pm 0.05$. The best fit values when allowing for non-flat geometries of the universe are given by the combination Las Campanas, peculiar velocities and supernovae; $h = 0.56_{-0.16}^{+0.21}$, $\Omega_\textrm{m} = 0.3_{-0.06}^{+0.09}$ and $\sigma_8 = 0.87 \pm 0.07$. For that combination, $\Omega_{\Lambda} = 0.78_{-0.12}^{+0.05}$ and $\Omega_{\textrm{k}} = -0.14_{-0.05}^{+0.17}$. For convenience, both $\Omega_{\Lambda}$ and $\Omega_{\textrm{k}}$ are marginalised over in the following pages. Figs. \ref{lss-histograms-1} and \ref{lss-histograms-2} show the marginalised one-dimensional posterior distribution of the Hubble parameter $h$, the matter density $\Omega_{\textrm m}$ and the mass density fluctuation parameter $\sigma_8$ calculated by the MCMC routine. 

Peculiar velocity data seem to prefer higher values of $\sigma_8$ (panel 3C of figure \ref{lss-histograms-1}) such that when combined with supernovae data, $\sigma_8 > 1$ at the 1$\sigma$ confidence level (panel 3C of figure \ref{lss-histograms-2}).

The combination of the peculiar velocities and supernovae data  provides $h > 0.6$ at 1$\sigma$. The combination of IRAS with peculiar velocities also prefers $0.55 \lesssim h$ whereas all other combinations prefer lower values of the Hubble parameter. The constraints on $h$ obtained here are wide compared to other estimates, e.g. \cite{HST-key}. Table 3 of \cite{wmap-spergel} shows that the latest determinations of $h$ are all in the range preferred here: $0.4 \leq h \leq 0.77$ at 1$\sigma$ in the full joint analysis.

In the combination of Las Campanas and peculiar velocities likelihood functions (panels 1B, 3B and 6B of Figure \ref{lss-histograms-1} respectively), $\Omega_m$ becomes less tightly constrained compared to the individual data sets and shifts to higher values. Note that the 1$\sigma$ intervals for $\Omega_{\textrm{m}}$ all overlap and the resulting interval from the combination of all data sets $\Omega_\textrm{m} = 0.28 \pm 0.05$ is almost identical with the latest results from \cite{wmap-spergel}, $\Omega_\textrm{m} = 0.29 \pm 0.07$.This demonstrates that owing to the non-linear relation between cosmological parameters, the resulting probability distribution function of derived parameters is not simply the product of the distributions obtained individually from each data set. 

Two trends appear in these results that seem related to whether the Las Campanas data are used. The combinations that include the Las Campanas data tend to favour lower values of the Hubble parameter and of the mass density fluctuation parameter. On the other hand, when IRAS or peculiar velocities data are used higher values of the Hubble parameter are preferred, more in accordance with the HST key project \cite{HST-key} or WMAP \cite{wmap-spergel} results. This is at the expense of a higher value of $\sigma_8$ as shown in panel 6C of Figure \ref{lss-histograms-2}.

\begin{table}[!bht]
\begin{center}
{\large
\begin{tabular}{|l|c|c|c|}\hline
data sets & $h$ & $\Omega_\textrm{m}$ & $\sigma_8$ \\ \hline
L & -- & $< 0.11$ & $0.91_{-0.08}^{+0.07}$ \\
I & -- & $0.13_{-0.08}^{+0.11}$ & $> 0.85$ \\
P & -- & $0.20_{-0.04}^{+0.16}$ & $1.48_{-0.44}$ \\
S & -- & $0.28_{-0.15}^{+0.1}$ & --\\
L + I & -- & $0.16_{0.06}^{0.13}$ & $0.87_{-0.06}^{+0.05}$ \\
L + P & $0.48_{-0.15}^{+0.2}$ & $0.38_{-0.14}^{+0.21}$ & $0.87_{-0.09}^{+0.06}$ \\
L + S & $> 0.34$ & $< 0.34$ & $0.87_{-0.07}^{+0.08}$ \\
I + P & $> 0.58$ & $0.2_{-0.07}^{+0.08}$ & $1.46_{-0.59}$ \\
I + S & $0.56_{-0.19}^{+0.4}$ & $0.25 \pm 0.06$ & $0.91_{-0.05}$ \\
P + S & $> 0.63$ & $0.28_{-0.11}^{+0.07}$ & $1.48_{-0.3}$\\
L + I + P & $> 0.33$ & $0.36_{-0.14}^{+0.16}$ & $0.85_{-0.06}^{+0.05}$\\
L + I + S & $0.63_{-0.23}^{+0.21}$ & $0.26\pm 0.05$ & $0.86_{-0.04}^{+0.05}$\\
L + P + S & $0.56_{-0.16}^{+0.21}$ & $0.3_{-0.06}^{+0.09}$ & $0.87 \pm 0.07$\\
I + P + S & $> 0.58$ & $0.25_{-0.05}^{+0.06}$ & $1.3_{-0.26}^{+0.39}$\\
L + I + P + S & $0.57_{-0.14}^{+0.16}$ & $0.28 \pm 0.05$ & $0.87_{-0.05}^{+0.04}$\\
\hline
\end{tabular}
}
\end{center}
\caption{Best-fit values for three cosmological parameters from different combinations of four data sets: Las Campanas (L), IRAS (I), Peculiar Velocities (P) and Supernovae (S).}
\label{lss-results-table}
\end{table}
 

\section{Conclusions}\label{lss-conclusions}

Four different cosmological sets of data provide constraints on  $h=0.57_{-0.14}^{+0.16}$, $\Omega_\textrm{m} = 0.28 \pm 0.05$ and $\sigma_8 = 0.87_{-0.05}^{+0.04}$ under the assumption of a flat universe, and $h = 0.56_{-0.16}^{+0.21}$, $\Omega_\textrm{m} = 0.3_{-0.06}^{+0.09}$ and $\sigma_8 = 0.87 \pm 0.07$ for all geometries. The full combination of these data sets provides remarkably tight constraints on $\Omega_\textrm{m}$, $h$ and $\sigma_8$. The resulting values of $/Omega_m$ and $\sigma_8$ are in excellent agreement with parameter estimation presented in \cite{wmap-spergel}:  $\Omega_m=0.27\pm0.04$, $\sigma_8=0.84 \pm 0.04$ although the constraint on the Hubble parameter, $h=0.71_{-0.04}^{+0.03}$, differs somewhat.

In general, the three parameters display a different behaviour depending on the data and priors. The Hubble parameter is ill-determined in each case and the tightest contraint is consistent at 1$\sigma$ with most $h$ estimates (see \cite{HST-key} and references therein). The matter density $\Omega_m$ is fairly well determined in each analysis and all results are in good agreement independently of which data are included or which priors are assumed. The amplitude of mass fluctutations parameter $\sigma_8$ is more difficult to constrain. As in the litterature (see e.g. table 4 in \cite{wmap-spergel}), several values have been derived that are not necessarily consistent and depend strongly on which data are used. Unlike the Hubble parameter which is allowed a wide range of values whatever the analysis, observations lead to rather tight constraints on $\sigma_8$ but different ones, thereby suggesting that the discrepancy is due to errors or misestimates in the data rather than to priors.

\subsubsection*{Acknowledgments}
The authors are grateful to Steve Gull and John Skilling for the {\scshape Bayesys} code. CJO acknowledges a Marie Curie Intra-European Fellowship grant \#501007.

\bibliographystyle{mn2e}

\end{document}